\documentclass[a4paper,12pt]{article}

\usepackage[margin=1in,dvips,a4paper,nohead]{geometry}
\usepackage[margin=1em, font=small, justification=centerlast,
tableposition=top]{caption}

\usepackage{amsfonts,amssymb,amsmath}
\usepackage[noBBpl,slantedGreek]{mathpazo}

\usepackage{pstricks}

\psset{unit=1pt, dimen=middle, linewidth=.5pt, arrowsize=3pt 2}

\newdimen \psx
\newdimen \psy

\def\psddots (#1,#2){
  \psx #1\psxunit
  \psy #2\psyunit
  \qdisk(\psx,\psy){.7pt}
  \advance \psx -.2\psxunit
  \advance \psy .2\psyunit
  \qdisk(\psx,\psy){.7pt}
  \advance \psx .4\psxunit
  \advance \psy -.4\psyunit
  \qdisk(\psx,\psy){.7pt}
}

\def\pssddots (#1,#2){
  \psx #1\psxunit
  \psy #2\psyunit
  \qdisk(\psx,\psy){.7pt}
  \advance \psx .2\psxunit
  \advance \psy .2\psyunit
  \qdisk(\psx,\psy){.7pt}
  \advance \psx -.4\psxunit
  \advance \psy -.4\psyunit
  \qdisk(\psx,\psy){.7pt}
}

\makeatletter
\def\numberbysection{\@addtoreset{equation}{section}
        \def\theequation{\thesection.\arabic{equation}}}

\numberbysection

\def \:{\mskip .5\thinmuskip}
\def\ph {{\hbox to 0pt{\phantom{$\scriptstyle -1$}\hss}}}

\def\bbC {\mathbb C}

\def\bbR {\mathbb R}
\def\bbZ {\mathbb Z}
\def\calA {\mathcal A}

\def\calF {\mathcal F}
\def\calG {\mathcal G}

\def\calL {\mathcal L}
\def\calM {\mathcal M}
\def\calN {\mathcal N}
\def\calO {\mathcal O}
\def\calP {\mathcal P}

\def\calR {\mathcal R}
\mathchardef\Gamma "100
\def\gothg{\mathfrak g}
\def\gothG{\mathfrak G}
\def\gothgl{\mathfrak{gl}}

\def\d {\mathrm{d}}
\def\rme {\mathrm e}

\def\rmi {\mathrm i}
\def\rmGL {\mathrm{GL}}

\def\wt {\widetilde}
\def\Mat {\mathrm{Mat}}

\title{\bf Solving non-abelian loop Toda equations}

\author{Kh. S. Nirov\thanks{On leave of absence from the
\small \em Institute for Nuclear Research of the Russian Academy of
Sciences, 60th October Ave 7a, 117312 Moscow,
Russia}~~ and~ A. V. Razumov\thanks{On leave of absence from the
\small \em Institute for High Energy Physics, 142281 Protvino, 
Moscow Region, Russia}\\
\small Fachbereich C--Physik, Bergische Universit\"at Wuppertal\\ 
\small D-42097 Wuppertal, Germany
}

\date{}

\begin{document}

\maketitle

\begin{abstract}
We construct soliton solutions for non-abelian loop Toda equations
associated with general linear groups. Here we consider the untwisted
case only and use the rational dressing method based upon appropriate
block-matrix representation suggested by the initial $\bbZ$-gradation.
\end{abstract}

\section{Introduction}

The Toda systems \cite{LezSav92,RazSav97,MikOlsPer81} associated
with loop groups possess features attractive from both mathematical
and physical perspectives. The fact that they have the so-called
soliton solutions is certainly among such interesting properties. 
Here, by an $N$-soliton solution we simply mean a solution depending 
on $N$ linear combinations of independent variables. In particular, 
the investigation of soliton solutions imply developing methods of 
solving nonlinear partial differential equations and besides, also 
modeling various nonlinear phenomena in particle physics and field 
theory, see, for example, the paper \cite{BueFerRaz02} and references 
therein.

There are various approaches to constructing soliton solutions for
loop Toda systems. The best known and elaborated among them are,
probably, the rational dressing formalism \cite{ZakSha79,Mik81},
that is a version of the inverse scattering method, and the
Hirota's approach \cite{Hir04,Hol92,ConFerGomZim93,MacMcG92,
AraConFerGomZim93,ZhuCal93} based on an appropriate change of
the field variables. Also certain combinations of these two
methods prove to be quite efficient in the purpose of finding
soliton solutions of Toda equations \cite{BueFerRaz02,AssFer07}.
Besides, it is worth while mentioning generalizations of the
Leznov--Saveliev \cite{OliSavUnd93,OliTurUnd92,OliTurUnd93} and
the B\"{a}cklund--Darboux
\cite{ForGib80, MatSal91, LiaOliTur93, RogSch02, Zhou06}
methods that were employed at Toda systems.

In a recent paper \cite{NirRaz08a}, we have carried out a comparative
analysis of the Hirota's and rational dressing methods in application
to abelian Toda systems associated with the untwisted loop groups of
complex general linear groups and, in particular, explicitly
reproduced the corresponding multi-soliton solutions. Further, in
the subsequent paper \cite{NirRaz08b}, we have constructed soliton
solutions for abelian twisted loop Toda systems. And now, we are
going to investigate the non-abelian loop Toda equations being a
direct generalization of the systems considered in \cite{NirRaz08a}.
Here we work within the rational dressing formalism based upon
appropriate block-matrix representation. The latter is naturally
suggested by the $\bbZ$-gradation under consideration and turns
out to be most suitable to the non-abelian Toda systems.

Inasmuch as the abelian soliton solutions allow for such a physical
interpretation as of interacting extended particle-like objects, so
their non-abelian generalizations should be very interesting as such
objects having additionally certain internal structures. And since
this physical interpretation promises a good basis for a consistent
modeling of various nonlinear phenomena, the mathematical part
consisting in developing the corresponding integration methods
and constructing explicit soliton solutions becomes crucial.

Note finally that since the pioneering paper \cite{Mik81} where
simplest non-abelian loop Toda equations were presented, certain
efforts have been made to solve them by means of various methods.
Thus, in \cite{EtiGelRet97a} the notion of quasi-determinants
was exploited for the purpose, see also \cite{XLiNim07}; in
\cite{ParShi95} a simplest matrix generalization of the sine-Gordon
equation was treated by the rational dressing method.\footnote{However,
it is not quite clear how the soliton solutions of \cite{ParShi95}
were obtained without averaging over the action of the corresponding
automorphism group, which is one of the principal ingredients of
the rational dressing procedure.} An approach based on the dressing
(gauge) transformation method was developed in a series of papers
\cite{GomGueSotZim01, CabGomGueSotZim01, GomGueSotZim00, GomZimSot99}
for a simplest case of non-abelian affine Toda systems where a
specific gradation leads to a minimal extension of the abelian
counterpart, and then the vertex operator method was also used
there in order to construct some soliton solutions.

\section{Formulation of loop Toda equations}

The formulation of Toda systems, in a way most appropriate to
our purposes, is based on their simple differential-geometry
and group-algebraic background, and here we generally follow
the monographs \cite{LezSav92, RazSav97} and the papers
\cite{NirRaz06, NirRaz07a, NirRaz07b}.

Let the trivial fiber bundle $\bbR^2 \times \calG \rightarrow \bbR^2$,
with the structure Lie group $\calG$ and its Lie algebra $\gothG$, be
given. We identify a connection in this fiber bundle with a
$\gothG$-valued $1$-form $\calO$ on $\bbR^2$ and decompose it
over basis $1$-forms,
\[
\calO = \calO_- {\mathrm{d}} z^- + \calO_+ {\mathrm{d}} z^+,
\]
where $z^-$, $z^+$ are the standard coordinates on the base manifold
$\bbR^2$, and the components $\calO_-$, $\calO_+$ are $\gothG$-valued
functions on it. We assume that the connection $\calO$ is flat, and
it means that its curvature is zero. Then, in terms of the components,
we have
\begin{equation}
\partial_- \calO_+ - \partial_+ \calO_- + [\calO_-, \calO_+] = 0,
\label{e:2.1}
\end{equation}
where we use the notation $\partial_- = \partial/\partial z^-$
and $\partial_+ = \partial/\partial z^+$. One can consider this
relation as a system of partial differential equations. The
general solution of this system is well known,
\[
\calO_- = \Phi^{-1} \partial_- \Phi, \qquad
\calO_+ = \Phi^{-1} \partial_+ \Phi,
\]
where $\Phi$ is an arbitrary mapping of $\bbR^2$ to $\calG$.
Actually, the zero-cur\-va\-tu\-re condition as a system of
partial differential equations is trivial due to the
gauge invariance. Indeed, if a connection $\calO$ satisfies
(\ref{e:2.1}), then for an arbitrary mapping $\Psi$ of
$\bbR^2$ to $\calG$ the gauge-transformed connection
\begin{equation}
\calO^\Psi = \Psi^{-1} \calO \Psi + \Psi^{-1} \d \Psi,
\label{e:2.2}
\end{equation}
satisfies (\ref{e:2.1}) as well.

To obtain nontrivial integrable systems out of the zero-curvature
condition we impose on the connection $\calO$ some restrictions
which destroy the gauge invariance. To come specifically to
Toda systems, we should use certain grading and gauge-fixing
conditions.

Suppose that $\gothG$ is endowed with a $\bbZ$-gradation,
\[
\gothG = \bigoplus_{k \in \bbZ} \gothG_k,
\qquad
[\gothG_k , \gothG_l] \subset \gothG_{k+l},
\]
and $L$ is such a positive integer that the grading subspaces
$\gothG_{k}$ and $\gothG_{-k}$, where $0 < k < L$, are trivial. The
grading condition states that the components of $\calO$ have the form
\begin{equation}
\calO_- = \calO_{-0} + \calO_{-L}, \qquad \calO_+ = \calO_{+0} +
\calO_{+L}, \label{e:2.3}
\end{equation}
where $\calO_{-0}$ and $\calO_{+0}$ take values in $\gothG_0$, while
$\calO_{-L}$ and $\calO_{+L}$ take values in $\gothG_{-L}$ and
$\gothG_{+L}$ respectively. There is a residual gauge invariance.
Indeed, the gauge transformation (\ref{e:2.2}) with $\Psi$ taking
values in the connected Lie subgroup $\calG_0$ corresponding to the
subalgebra $\gothG_0$ does not violate the grading condition
(\ref{e:2.3}). Therefore, we additionally impose a
gauge-fixing condition of the form
\[
\calO_{+0} = 0.
\]
Now the components of the connection $\calO$ can be represented as
\begin{equation}
\calO_- = \Xi^{-1} \partial_- \Xi + \calF_-, \qquad \calO_+ = \Xi^{-1}
\calF_+ \Xi,
\label{e:2.4}
\end{equation}
where $\Xi$ is a mapping of $\bbR^2$ to $\calG_0$, $\calF_-$ and
$\calF_+$ are some mappings of $\bbR^2$ to $\gothG_{-L}$ and
$\gothG_{+L}$. One can easily see that the zero-curvature
condition is equivalent to the equality\footnote{We assume for
simplicity that $\calG$ is a subgroup of the group formed by
invertible elements of some unital associative algebra $\calA$.
In this case $\gothG$ can be considered as a subalgebra of the
Lie algebra associated with $\calA$. Our consideration can be
generalized to the case of an arbitrary Lie group~$\calG$.}
\begin{equation}
\partial_+ (\Xi^{-1} \partial_- \Xi) = [\calF_-, \Xi^{-1} \calF_+ \Xi]
\label{e:2.5}
\end{equation}
and the relations
\begin{equation}
\partial_+ \calF_- = 0, \qquad \partial_- \calF_+ = 0.
\label{e:2.6}
\end{equation}
We suppose that the mappings $\calF_-$ and $\calF_+$ are fixed and
consider (\ref{e:2.5}) as an equation for $\Xi$ called the Toda
equation. When the group $\calG_0$ is abelian the corresponding Toda
equations are called abelian. In other cases we have non-abelian
Toda systems.

Thus, a Toda equation associated with a Lie group $\calG$ is specified
by a choice of a $\bbZ$-gradation of the Lie algebra $\gothG$ of
$\calG$ and mappings $\calF_-$, $\calF_+$ satisfying the conditions
(\ref{e:2.6}). To classify the Toda equations associated with a
Lie group $\calG$ one should classify the $\bbZ$-gradations of
the Lie algebra $\gothG$ of $\calG$.

We consider the case where $\calG$ is a loop group of a
finite-dimensional Lie group, $\calL_{a,M}(G)$, where $a$ is an
automorphism of $G$ of order $M$. The corresponding Lie algebra
$\gothG$ is thus the loop Lie algebra $\calL_{A,M}(\gothg)$, where
$\gothg$ is the Lie algebra of the Lie group $G$, with $A$ being the
respective automorphism of $\gothg$ of order $M$. This is a subalgebra
of the loop Lie algebra $\calL(\gothg)$ formed by elements $\xi$
satisfying the equality
\[
\xi(\epsilon_M s) = A(\xi(s)),
\]
where $\epsilon_M = \rme^{2 \pi \rmi/M}$ is the $M$th principal root
of unity and $s \in S^1$. Similarly, the loop group $\calL_{a,M}(G)$
is defined as the subgroup of the loop group $\calL(G)$ formed by
the elements $\chi$ satisfying the equality
\[
\chi(\epsilon_M s) = a(\chi(s)).
\]
For a consistent description of these objects given in a way
most suitable for the matter of loop Toda equations we refer
to \cite{NirRaz06,NirRaz07a,NirRaz07b}. To have a wider list
of publications on such systems see also
\cite{FerMirGui95,FerMirGui97,FerGalHolMir97,Mir99}
and references therein.

To specify the Toda equations associated with the loop group
$\calL_{a,M}(G)$ we first note that the group $\calL_{a,M}(G)$
and its Lie algebra $\calL_{A,M}(\gothg)$ are infinite-dimensional
manifolds. However, using the so-called exponential law
\cite{KriMic91,KriMic97}, that can generally
be expressed by the canonical identification
\[
C^\infty(\calM,C^\infty(\calN,\calP)) =
C^\infty(\calM \times \calN,\calP),
\]
where $\calM$, $\calN$, $\calP$ are finite-dimensional manifolds
and $\calN$ is besides compact, we reformulate the zero-curvature
representation of the Toda equations associated with $\calL_{a,M}(G)$
in terms of finite-dimensional manifolds.

In the case under consideration the connection components $\calO_-$
and $\calO_+$ entering the equality (\ref{e:2.1}) are mappings of
$\bbR^2$ to the loop Lie algebra $\calL_{A,M}(\gothg)$. We denote
the corresponding mappings of $\bbR^2 \times S^1$ to $\gothg$ by
$\omega_-$ and $\omega_+$, and call them also the connection
components. The mapping $\Phi$ generating the connection is
a mapping of $\bbR^2$ to $\calL_{a,M}(G)$. Denoting the
corresponding mapping of $\bbR^2 \times S^1$ by $\varphi$
we write
\begin{equation}
\varphi^{-1} \partial_- \varphi = \omega_-, \qquad \varphi^{-1}
\partial_+ \varphi = \omega_+. \label{e:2.7}
\end{equation}
Seeing that the mapping $\varphi$ uniquely determines the mapping
$\Phi$, we say that the mapping $\varphi$ also generates the
connection under consideration.

We follow the classification of loop Toda systems performed in
\cite{NirRaz06,NirRaz07a,NirRaz07b}. Important for our purposes
here is that the initial Toda equation associated with
$\calL_{a,M}(G)$ is equivalent to a Toda equation associated
with $\calL_{a',M'}(G)$ arising when $\calL_{A',M'}(\gothg)$
is supplied with the standard $\bbZ$-gradation.

The grading subspaces for the standard $\bbZ$-gradation
of a loop Lie algebra $\calL_{A,M}(\gothg)$ are
\[
\calL_{A, M}(\gothg)_{k} = \{ \xi \in \calL_{A,M}(\gothg)
\mid \xi = \lambda^k x, \ A(x) = \epsilon_M^{k} x \},
\]
where by $\lambda$ we denote the restriction of the standard
coordinate on $\bbC$ to $S^1$. It is clear that every automorphism
$A$ of the Lie algebra $\gothg$ satisfying the relation
$A^M = \mathrm{id}_\gothg$ induces a $\mathbb Z_M$-gradation
of $\gothg$ with the grading subspaces
\[
\gothg_{[k]_M} = \{x \in \gothg \mid A(x) = \epsilon_M^{k} x\},
\qquad k = 0, \ldots, M-1,
\]
where by $[k]_M$ we denote the element of the ring
$\mathbb Z_M$ corresponding to the integer $k$. Vice versa,
any $\mathbb Z_M$-gradation of $\gothg$ obviously defines
an automorphism $A$ of $\gothg$ satisfying the relation
$A^M = \mathrm{id}_\gothg$. In terms of the corresponding
$\bbZ_M$-gradation the grading subspaces for the standard
$\bbZ$-gradation of a loop Lie algebra $\calL_{A,M}(\gothg)$
are
\[
\calL_{A, M}(\gothg)_{k} = \{ \xi \in \calL_{A,M}(\gothg)
\mid \xi = \lambda^k x, \ x \in \gothg_{[k]_M} \}.
\]

It is evident that for the standard $\bbZ$-gradation the subalgebra
$\calL_{A,M}(\gothg)_0$ is isomorphic to the subalgebra
$\gothg_{[0]_M}$ of $\gothg$, and the Lie group $\calL_{a,M}(G)_0$
is isomorphic to the connected Lie subgroup $G_0$ of $G$ corresponding
to the Lie algebra $\gothg_{[0]_M}$. Hence, the relations (\ref{e:2.4})
are equivalent to the relations
\begin{equation}
\omega_- = \gamma^{-1} \partial_- \gamma + \lambda^{-L} c_-,
\qquad
\omega_+ = \lambda^L \gamma^{-1} c_+ \gamma, \label{e:2.8}
\end{equation}
where $\gamma$, taken as a smooth mapping of $\bbR^2 \times S^1$
to $G$ corresponding to the mapping $\Xi$ in accordance with the
exponential law, is actually a mapping of $\bbR^2$ to $G_0$, and
respecting the mappings $\calF_-$ and $\calF_+$, the mappings
$c_-$ and $c_+$ above are mappings of $\bbR^2$ to $\gothg_{-[L]_M}$
and $\gothg_{+[L]_M}$ respectively. The Toda equation can
subsequently be written as
\begin{equation}
\partial_+(\gamma^{-1} \partial_- \gamma) = [c_-, \gamma^{-1} c_+
\gamma]. \label{e:2.9}
\end{equation}
The conditions (\ref{e:2.6}) imply that
\begin{equation}
\partial_+ c_- = 0, \qquad \partial_- c_+ = 0. \label{e:2.10}
\end{equation}
We call an equation of the form (\ref{e:2.9}) also a Toda equation.

Let us consider the transformations
\begin{gather}
\gamma' = \eta_+^{-1} \: \gamma \: \eta_-, \label{e:2.11} \\
c'_- = \eta_-^{-1} c_- \eta_-, \qquad c'_+
= \eta_+^{-1} c_+ \eta^{}_+, \label{e:2.12}
\end{gather}
where $\eta_-$ and $\eta_+$ are some mappings of $\bbR^2 \times S^1$
to $G_0$ that satisfy the conditions
\[
\partial_+ \eta_- = 0, \qquad \partial_- \eta_+ = 0.
\]
If a mapping $\gamma$ satisfies the Toda equation (\ref{e:2.9}), then
the mapping $\gamma'$ satisfies the Toda equation (\ref{e:2.9}) where
the mappings $c_-$, $c_+$ are replaced by the mappings $c'_-$ and
$c'_+$. If the mappings $\eta_-$ and $\eta_+$ are such that
\[
\eta_-^{-1} c_- \eta_- = c_-, \qquad \eta_+^{-1} c_+ \eta^{}_+ = c_+
\]
then the transformation (\ref{e:2.11}) is a symmetry transformation
for the Toda equation under consideration.

\section{Untwisted loop Toda equations} \label{s:3}

The complete classification of Toda equations associated with
twisted loop groups of complex classical Lie groups, where
the corresponding twisted loop Lie algebras are endowed with
integrable $\bbZ$-gradations with finite-dimensional grading
subspaces, is given in the series of papers \cite{NirRaz06,
NirRaz07a, NirRaz07b}. We will use these results related to
the particular case of untwisted loop groups of the complex
general linear groups. The $\bbZ$-gradations of the corresponding
loop Lie algebras are thus generated by an inner automorphism
of the initial finite-dimensional complex Lie algebra
$\gothgl_n(\bbC)$,
\[
A(x) = h \: x \: h^{-1},
\]
where $x$ is an arbitrary element of $\gothgl_n(\bbC)$, and $h$ is
a diagonal matrix of the form
\[
h = \left( \begin{array}{cccc}
\epsilon_M^{m_1} I_{n_1} & & & \\
& \epsilon_M^{m_2} I_{n_2} & & \\
& & \ddots & \\
& & & \epsilon_M^{m_p} I_{n_p}
\end{array} \right),
\]
where $I_{n_\alpha}$ denotes the $n_\alpha \times n_\alpha$ unit matrix
and $M \ge m_1 > m_2 > \ldots > m_p > 0$. Here $n_\alpha$,
$\alpha = 1,\ldots,p$, are positive integers, such that
$\sum_{\alpha=1}^p n_\alpha = n$. According to the block-matrix structure
of $h$, it is convenient to represent the element $x$ as a $p \times p$
block matrix $(x_{\alpha\beta})$, where $x_{\alpha \beta}$ is an
$n_\alpha \times n_\beta$ matrix,
\begin{equation}
x = \left( \begin{array}{cccc}
x_{11} & x_{12} &\ldots& x_{1p} \\
x_{21} & x_{22} &\ldots& x_{2p} \\
\vdots & \vdots & \ddots & \vdots \\
x_{p1} & x_{p2} &\ldots& x_{pp}
\end{array} \right).
\label{e:3.1}
\end{equation}
Here the inner automorphism $a$ acts on an arbitrary element $g$ of
$\rmGL_n(\bbC)$ as $a(g) = h g h^{-1}$, with the same diagonal matrix
$h$ given above.

The mapping $\gamma$ has the block-diagonal form
\[
\psset{xunit=1.7em, yunit=1.2em}
\gamma = \left( \raise -1.8\psyunit \hbox{\begin{pspicture}(.6,.6)
(4.5,4.2)
\rput(1,4){$\Gamma_1$}
\rput(2,3){$\Gamma_2$}
\qdisk(2.7,2.3){.7pt} \qdisk(3,2){.7pt} \qdisk(3.3,1.7){.7pt}
\rput(4,1){$\Gamma_p$}
\end{pspicture}} \right).
\]
For each $\alpha = 1, \ldots, p$ the mapping $\Gamma_\alpha$ is a
mapping of $\bbR^2$ to the Lie group $\rmGL_{n_\alpha}(\bbC)$.

The mapping $c_+$ has the following block-matrix structure:
\begin{equation}
\psset{xunit=2.5em, yunit=1.4em}
c_+ = \left( \raise -2.4\psyunit \hbox{\begin{pspicture}(.6,.5)
(5.6,5.3)
\rput(1,5){$0$} \rput(2,4.92){$C_{+1}$}
\rput(2,4){$0$}
\qdisk(2.8,4.2){.7pt} \qdisk(3,4){.7pt} \qdisk(3.2,3.8){.7pt}
\qdisk(2.8,3.2){.7pt} \qdisk(3,3){.7pt} \qdisk(3.2,2.8){.7pt}
\qdisk(3.8,3.2){.7pt} \qdisk(4,3){.7pt} \qdisk(4.2,2.8){.7pt}
\rput(4,2){$0$} \rput(5,1.87){$C_{+(p-1)}$}
\rput(1,.94){$C_{+0}$} \rput(5,1){$0$}
\end{pspicture}} \right),
\label{e:3.2}
\end{equation}
where for each $\alpha = 1, \ldots, p-1$ the mapping $C_{+\alpha}$ is
a mapping of $\bbR^2$ to the space of $n_\alpha \times n_{\alpha+1}$
complex matrices, and $C_{+0}$ is a mapping of $\bbR^2$ to the space
of $n_p \times n_1$ complex matrices. The mapping $c_-$ has a similar
block-matrix structure:
\begin{equation}
\psset{xunit=2.5em, yunit=1.4em}
c_- = \left( \raise -2.4\psyunit \hbox{\begin{pspicture}(.5,.5)
(5.5,5.3)
\rput(1,5){$0$} \rput(5,4.92){$C_{-0}$}
\rput(1,4){$C_{-1}$} \rput(2,4){$0$}
\qdisk(1.8,3.2){.7pt} \qdisk(2,3){.7pt} \qdisk(2.2,2.8){.7pt}
\qdisk(2.8,3.2){.7pt} \qdisk(3,3){.7pt} \qdisk(3.2,2.8){.7pt}
\qdisk(2.8,2.2){.7pt} \qdisk(3,2){.7pt} \qdisk(3.2,1.8){.7pt}
\rput(4,1.87){$0$}
\rput(4,.94){$C_{-(p-1)}$} \rput(5,1){$0$}
\end{pspicture}} \right),
\label{e:3.3}
\end{equation}
where for each $\alpha = 1, \ldots, p-1$ the mapping $C_{-\alpha}$ is
a mapping of $\bbR^2$ to the space of $n_{\alpha+1} \times n_\alpha$
complex matrices, and $C_{-0}$ is a mapping of $\bbR^2$ to the space
of $n_1 \times n_p$ complex matrices. The conditions (\ref{e:2.10})
imply
\[
\partial_+ C_{-\alpha} = 0, \qquad \partial_- C_{+\alpha} = 0, \qquad
\alpha = 0, 1, \ldots, p-1.
\]

It is not difficult to show that the Toda equation (\ref{e:2.9}) is
equivalent to the following system of equations for the mappings
$\Gamma_\alpha$:
\begin{align}
\partial_+ \left( \Gamma_1^{-1} \: \partial_- \Gamma_1^{} \right)
&= - \Gamma_1^{-1} C_{+1}^{} \: \Gamma_2^{} \: C_{-1}^{}
+ C_{-0}^{} \Gamma_p^{-1} C_{+0}^{} \Gamma_1^{},
\notag \\*
\partial_+ \left( \Gamma_2^{-1} \: \partial_- \Gamma_2^{} \right)
&= - \Gamma_2^{-1} C_{+2}^{} \: \Gamma_3^{} \: C_{-2}^{}
+ C_{-1}^{} \Gamma_1^{-1} C_{+1}^{} \Gamma_2^{},
\notag \\*
& \quad \vdots
\label{e:3.4} \\*
\partial_+ \left(\Gamma_{p-1}^{-1} \: \partial_-
\Gamma_{p-1}^{}\right)
&= - \Gamma_{p-1}^{-1} C_{+(p-1)}^{} \: \Gamma_p^{} \: C_{-(p-1)}^{}
+ C_{-(p-2)}^{} \Gamma_{p-2}^{-1} C_{+(p-2)}^{} \Gamma_{p-1}^{},
\notag \\*
\partial_+ \left( \Gamma_p^{-1} \: \partial_- \Gamma_p^{} \right)
&= - \Gamma_p^{-1} C_{+0}^{} \: \Gamma_1^{} \: C_{-0}^{}
+ C_{-(p-1)}^{} \Gamma_{p-1}^{-1} C_{+(p-1)}^{} \Gamma_p^{}.
\notag
\end{align}
As is shown in \cite{NirRaz07a, NirRaz07b}, if for some $\alpha$
we have $C_{-\alpha} = 0$ or $C_{+\alpha} = 0$, then the system of
equations (\ref{e:3.4}) is equivalent to a system of equations
associated with a respective finite-dimensional Lie group, or
to a set of two such systems. Hence, to deal actually with
Toda equations associated with a loop group, we assume that
all mappings $C_{-\alpha}$ and $C_{+\alpha}$ are nontrivial.
This is possible only if $m_\alpha = (p - \alpha + 1)L$
and $M = p L$. Moreover, it appears that in the case
under consideration we can assume, without any loss
of generality, that the positive integer $L$ is equal
to $1$.

The Toda equations (\ref{e:3.4}) can also be written as
\begin{equation}
\partial_+ (\Gamma^{-1}_\alpha \partial_- \Gamma^{}_\alpha)
+ \Gamma^{-1}_\alpha C_{+\alpha} \Gamma^{}_{\alpha+1} C_{-\alpha}
- C_{-(\alpha-1)} \Gamma^{-1}_{\alpha-1} C_{+(\alpha-1)}
\Gamma_{\alpha} = 0,
\label{e:3.5}
\end{equation}
with $\Gamma_\alpha$ subject to the periodicity condition
$\Gamma_{\alpha+p} = \Gamma_\alpha$. If transformed according
to (\ref{e:2.11}), (\ref{e:2.12}), the submatrices entering
the Toda equations would look here as follows:
\begin{equation}
\Gamma'_\alpha = \eta^{-1}_{+\alpha} \Gamma_\alpha \eta_{-\alpha},
\qquad
C'_{-\alpha} = \eta^{-1}_{-(\alpha+1)} C_{-\alpha} \eta_{-\alpha},
\qquad
C'_{+\alpha} = \eta^{-1}_{+\alpha} C_{+\alpha} \eta_{+(\alpha+1)},
\label{e:3.6}
\end{equation}
with the block-diagonal matrices $\eta_\pm$ defined by
$(\eta_\pm)_{\alpha\beta} = \eta_{\pm \alpha} \delta_{\alpha\beta}$.

Similarly to the abelian case \cite{NirRaz08a}, it can be shown
that the determinant of the mapping $\gamma$ can be represented
in a factorized form as
\[
\det \gamma = \prod_{\alpha = 1}^p \det \Gamma_\alpha
= \Gamma_+^{} \Gamma_-^{-1},
\]
where
\[
\partial_+ \Gamma_- = 0, \qquad \partial_- \Gamma_+ = 0.
\]
Then, setting
\[
\eta_{-\alpha} = \Gamma_-^{1/n} I_{n_\alpha}, \qquad
\eta_{+\alpha} = \Gamma_+^{1/n} I_{n_\alpha}
\]
in (\ref{e:3.6}), we can see that it is possible to make
the determinant of the transformed mapping $\gamma'$ be equal
to $1$,
\[
\det \gamma' = \prod_{\alpha = 1}^p \det \Gamma'_\alpha
= 1.
\]
Therefore, the reduction to the non-abelian Toda systems associated
with the loop groups of the special linear groups is possible, just
as well as it was in the abelian case \cite{NirRaz08a}.

\section{Rational dressing} \label{s:4}

We require that for any $m \in \bbR^2$ the matrices $c_-(m)$
and $c_+(m)$ commute, that is equivalent to the relations
\begin{equation}
C_{-(\alpha-1)} C_{+(\alpha-1)} - C_{+\alpha} C_{-\alpha} = 0.
\label{e:4.1}
\end{equation}
Then it is obvious that
\begin{equation}
\gamma = I_n,
\label{e:4.2}
\end{equation}
where $I_n$ is the $n \times n$ unit matrix, is a solution to the
Toda equation (\ref{e:2.9}). Denote a mapping of $\bbR^2 \times S^1$
to $\rmGL_n(\bbC)$ which generates the corresponding connection by
$\varphi$. Using the equalities (\ref{e:2.7}) and (\ref{e:2.8})
and remembering that in our case $L = 1$, we write
\[
\varphi^{-1} \partial_- \varphi = \lambda^{-1} c_-, \qquad
\varphi^{-1} \partial_+ \varphi = \lambda \: c_+,
\]
where the matrices $c_+$ and $c_-$ are defined by the relations
(\ref{e:3.2}), (\ref{e:3.3}).

To construct more interesting solutions to the Toda equations we will
look for a mapping $\psi$, such that the mapping
\begin{equation}
\varphi' = \varphi \: \psi
\label{e:4.3}
\end{equation}
would generate a connection satisfying the grading condition and the
gauge-fixing constraint $\omega_{+0} = 0$.

For any $m \in \bbR^2$ the mapping $\tilde \psi_m$ defined by the
equality $\tilde \psi_m(s) = \psi(m, s)$, $s \in S^1$, is a smooth
mapping of $S^1$ to $\rmGL_n(\bbC)$. We treat the unit circle $S^1$ as
a subset of the complex plane which, in turn, is a subset of the
Riemann sphere. Assume that it is possible to extend
analytically each mapping $\tilde \psi_m$ to all of the Riemann
sphere. As the result we get a mapping of the direct product of
$\bbR^2$ and the Riemann sphere to $\rmGL_n(\bbC)$ which we also
denote by $\psi$. Suppose that for any $m \in \bbR^2$ the analytic
extension of $\tilde \psi_m$ results in a rational mapping regular at
the points $0$ and $\infty$, hence the name rational dressing. Below,
for each point $s$ of the Riemann sphere we denote by $\psi_s$ the
mapping of $\bbR^2$ to $\rmGL_n(\bbC)$ defined by the equality
$\psi_s(m) = \psi(m, s)$.

We work with the Toda equations described in section \ref{s:3}.
It means that the mapping $\psi$ is generated by a mapping of
the Euclidean plane to the loop group $\calL_{a,p}(\rmGL_n(\bbC))$
with the corresponding inner automorphism of order $p$. Hence,
for any $m \in \bbR^2$ and $s \in S^1$ we should have
\begin{equation}
\psi(m, \epsilon_p s) = h \: \psi(m, s) \: h^{-1},
\label{e:4.4}
\end{equation}
where $h$ is a block-diagonal matrix explicitly given by the expression
\begin{equation}
h_{\alpha,\beta} = \epsilon_p^{p-\alpha+1} I_{n_\alpha} \delta_{\alpha\beta},
\qquad \alpha, \beta = 1,2,\ldots,p.
\label{*}
\end{equation}
The equality (\ref{e:4.4}) means that two rational mappings
coincide on $S^1$, therefore, they must coincide on the entire
Riemann sphere.

A mapping, satisfying the equality (\ref{e:4.4}), can be constructed
by the following procedure. Let $\chi$ be an arbitrary mapping of the
direct product of $\bbR^2$ and the Riemann sphere to the algebra
$\Mat_n(\bbC)$ of $n \times n$ complex matrices. Let $\hat a$ be
a linear operator acting on $\chi$ as
\[
\hat a \: \chi (m, s) = h \: \chi(m, \epsilon_p^{-1} s) \: h^{-1}.
\]
It is easy to get convinced that the mapping
\[
\psi = \sum_{k=1}^p \hat a^k \chi
\]
satisfies the relation $\hat a \: \psi = \psi$ which is equivalent
to the equality (\ref{e:4.4}). Here we have $\hat a^p \chi = \chi$.
Note that $\chi$ is in fact a mapping to the Lie group
$\rmGL_n(\bbC)$, but, to justify the above averaging
relation, we should consider $\rmGL_n(\bbC)$ as a subset
of $\Mat_n(\bbC)$.

To construct a rational mapping satisfying (\ref{e:4.4}) we start
with a rational mapping regular at the points $0$ and $\infty$ and
having poles at $r$ different nonzero points $\mu_i$, $i = 1, \ldots,
r$. Concretely speaking, we consider a mapping $\chi$ of the form
\[
\chi = \left( I_n + p \sum_{i=1}^r
\frac{\lambda}{\lambda - \mu_i} P_i \right) \chi_0,
\]
where $P_i$ are some smooth mappings of $\bbR^2$ to the algebra
$\Mat_n(\bbC)$ and $\chi_0$ is a mapping of $\bbR^2$ to the Lie
subgroup of $\rmGL_n(\bbC)$ formed by the elements
$g \in \rmGL_n(\bbC)$ satisfying the equality
\begin{equation}
h g h^{-1} = g, \label{e:4.5}
\end{equation}
where $h$ is given by the expression (\ref{*}). Actually this
subgroup coincides with the subgroup $G_0$. The averaging procedure
leads to the mapping
\begin{equation}
\psi = \left(I_n + \sum_{i=1}^r \sum_{k=1}^p \frac{\lambda}
{\lambda - \epsilon_p^k \mu_i} h^k P_i \: h^{-k} \right) \psi_0,
\label{e:4.6}
\end{equation}
where $\psi_0 = p \chi_0$. It is convenient to assume that
$\mu_i^p \ne \mu_j^p$ for all $i \ne j$.

Denote by $\psi^{-1}$ the mapping of $\bbR^2 \times S^1$ to
$\rmGL_n(\bbC)$ defined by the relation
\[
\psi^{-1}(m, s) = (\psi(m, s))^{-1}.
\]
Suppose that for any fixed $m \in \bbR^2$ the mapping
$\tilde\psi_m^{-1}$ of $S^1$ to $\rmGL_n(\bbC)$ can be extended
analytically to a mapping of the Riemann sphere to $\rmGL_n(\bbC)$ and
as the result we obtain a rational mapping of the same structure as
the mapping $\psi$,
\begin{equation}
\psi^{-1} = \psi_0^{-1} \left( I_n
+ \sum_{i = 1}^r \sum_{k=1}^p \frac{\lambda}
{\lambda - \epsilon_p^k \nu_i} \: h^k \: Q_i \: h^{-k} \right),
\label{e:4.7}
\end{equation}
with the pole positions satisfying the conditions $\nu_i \ne 0$,
$\nu_i^p \ne \nu_j^p$ for all $i \ne j$, and additionally $\nu_i^p \ne
\mu_j^p$ for any $i$ and $j$. We will denote the mapping of the direct
product of $\bbR^2$ and the Riemann sphere to $\rmGL_n(\bbC)$ again by
$\psi^{-1}$.

By definition, the equality
\[
\psi^{-1} \psi = I_n
\]
is valid at all points of the direct product of $\bbR^2$ and $S^1$.
Since $\psi^{-1} \psi$ is a rational mapping, the above equality is
valid at all points of the direct product of $\bbR^2$ and the Riemann
sphere. Hence, the residues of $\psi^{-1} \psi$ at the points $\nu_i$
and $\mu_i$ should be equal to zero. Explicitly we have
\begin{gather}
Q_i \left( I_n + \sum_{j = 1}^r \sum_{k=1}^p \frac{\nu_i}{\nu_i -
\epsilon_p^k \: \mu_j} \: h^k \: P_j \: h^{-k} \right) = 0,
\label{e:4.8} \\
\left( I_n + \sum_{j = 1}^r \sum_{k=1}^p \frac{\mu_i}{\mu_i -
\epsilon_p^k \: \nu_j} \: h^k \: Q_j \: h^{-k} \right) P_i = 0.
\label{e:4.9}
\end{gather}
We will discuss later how to satisfy these relations, and now let us
consider what connection is generated by the mapping $\varphi'$
defined by (\ref{e:4.3}) with the mapping $\psi$ possessing the
prescribed properties.

Using the representation (\ref{e:4.3}), we obtain for the components
of the connection generated by $\varphi'$ the expressions
\begin{gather}
\omega_- = \psi^{-1} \partial_- \psi + \lambda^{-1} \psi^{-1} c_-
\psi, \label{e:4.10} \\
\omega_+ = \psi^{-1} \partial_+ \psi + \lambda \psi^{-1} c_+ \psi.
\label{e:4.11}
\end{gather}
We see that the component $\omega_-$ is a rational mapping which has
simple poles at the points $\mu_i$, $\nu_i$ and zero.\footnote{Here
and below discussing the holomorphic properties of mappings and
functions we assume that the point of the space $\bbR^2$ is arbitrary
but fixed.} Similarly, the component $\omega_+$ is a rational mapping
which has simple poles at the points $\mu_i$, $\nu_i$ and infinity.
We are looking for a connection which satisfies the grading and
gauge-fixing conditions. The grading condition in our case is the
requirement that for each point of $\bbR^2$ the component $\omega_-$
is rational and has the only simple pole at zero, while the component
$\omega_+$ is rational and has the only simple pole at infinity.
Hence, we demand that the residues of $\omega_-$ and $\omega_+$ at the
points $\mu_i$ and $\nu_i$ should vanish.

The residues of $\omega_-$ and $\omega_+$ at the points $\nu_i$ are
equal to zero if and only if
\begin{gather}
(\partial_- Q_i - \nu_i^{-1} Q_i \: c_-) \left( I_n + \sum_{j = 1}^r
\sum_{k=1}^p \frac{\nu_i}{\nu_i - \epsilon_p^k \: \mu_j} \: h^k \: P_j
\: h^{-k}
\right) = 0,
\label{e:4.12} \\
(\partial_+ Q_i - \nu_i \: Q_i \: c_+) \left( I_n + \sum_{j = 1}^r
\sum_{k=1}^p \frac{\nu_i}{\nu_i - \epsilon_p^k \: \mu_j} \: h^k \: P_j
\: h^{-k}
\right) = 0,
\label{e:4.13}
\end{gather}
respectively. Similarly, the requirement of vanishing of the residues
at the points $\mu_i$ gives the relations
\begin{gather}
\left( I_n + \sum_{j = 1}^r \sum_{k=1}^p \frac{\mu_i}{\mu_i -
\epsilon_p^k \: \nu_j} \: h^k \: Q_j \: h^{-k} \right) (\partial_- P_i +
\mu_i^{-1} c_- \: P_i) = 0,
\label{e:4.14} \\
\left( I_n + \sum_{j = 1}^r \sum_{k=1}^p \frac{\mu_i}{\mu_i -
\epsilon_p^k \: \nu_j} \: h^k \: Q_j \: h^{-k} \right)
(\partial_+ P_i + \mu_i \: c_+ \: P_i) = 0.
\label{e:4.15}
\end{gather}
To obtain the relations (\ref{e:4.12})--(\ref{e:4.15}) we made
use of the equalities (\ref{e:4.8}), (\ref{e:4.9}).

Suppose that we have succeeded in satisfying the relations
(\ref{e:4.8}), (\ref{e:4.9}) and (\ref{e:4.12})--(\ref{e:4.15}).
In such a case from the equalities (\ref{e:4.10}) and (\ref{e:4.11})
it follows that the connection under consideration satisfies the
grading condition.

It is easy to see from (\ref{e:4.11}) that
\[
\omega_+(m, 0) = \psi_0^{-1}(m) \partial_+ \psi_0(m).
\]
Taking into account that $\omega_{+0}(m) = \omega_+(m, 0)$, we
conclude that the gauge-fixing constraint $\omega_{+0} = 0$ is
equivalent to the relation
\begin{equation}
\partial_+ \psi_0 = 0. \label{e:4.16}
\end{equation}
Assuming that this relation is satisfied, we come to a connection
satisfying both the grading condition and the gauge-fixing condition.

Recall that if a flat connection $\omega$ satisfies the grading and
gauge-fixing conditions, then there exist a mapping $\gamma$ from
$\bbR^2$ to $G$ and mappings $c_-$ and $c_+$ of $\bbR^2$ to
$\gothg_{-1}$ and $\gothg_{+1}$, respectively, such that the
representation (\ref{e:2.8}) for the components $\omega_-$
and $\omega_+$ is valid. In general, the mappings $c_-$ and $c_+$
parameterizing the connection components may be different from the
mappings $c_-$ and $c_+$ which determine the mapping $\varphi$. Let us
denote the mappings corresponding to the connection under
consideration by $\gamma'$, $c_-'$ and $c_+'$. Thus, we have
\begin{align}
\psi^{-1} \partial_- \psi + \lambda^{-1} \psi^{-1} c_- \psi &=
\gamma^{\prime -1} \partial_- \gamma' + \lambda^{-1} c_-',
\label{e:4.17} \\
\psi^{-1} \partial_+ \psi + \lambda \psi^{-1} c_+ \psi &= \lambda
\gamma^{\prime -1} c_+' \gamma'. \label{e:4.18}
\end{align}
Note that $\psi_\infty$ is a mapping of $\bbR^2$ to the Lie subgroup
of $\rmGL_n(\bbC)$ defined by the relation (\ref{e:4.5}). Recall that
this subgroup coincides with $G_0$, and denote $\psi_\infty$ by
$\gamma$. {}From the relation (\ref{e:4.17}) we obtain the equality
\[
\gamma^{\prime -1} \partial_- \gamma' = \gamma^{-1} \partial_- \gamma.
\]
The same relation (\ref{e:4.17}) gives
\[
\psi^{-1}_0 c_- \psi_0 = c_-'.
\]
Impose the condition $\psi_0 = I_n$, which is consistent with
(\ref{e:4.16}). Here we have
\[
c_-' = c_-.
\]
Finally, from (\ref{e:4.18}) we obtain
\[
\gamma^{\prime -1} c_+' \gamma' = \gamma^{-1} c_+ \gamma.
\]
We see that if we impose the condition $\psi_0 = I_n,$ then the
components of the connection under consideration have the form
given by (\ref{e:2.8}) where $\gamma = \psi_\infty$.

Thus, to find solutions to Toda equations under consideration, we can
use the following procedure. Fix $2r$ complex numbers $\mu_i$ and
$\nu_i$. Find matrix-valued functions $P_i$ and $Q_i$ satisfying
the relations (\ref{e:4.8}), (\ref{e:4.9}) and
(\ref{e:4.12})--(\ref{e:4.15}). With the help
of (\ref{e:4.6}), (\ref{e:4.7}), assuming that
\[
\psi_0 = I_n,
\]
construct the mappings $\psi$ and $\psi^{-1}$. Then, the mapping
\begin{equation}
\gamma = \psi_\infty
\label{e:4.19}
\end{equation}
satisfies the Toda equation (\ref{e:2.9}).

Let us return to the relations (\ref{e:4.8}), (\ref{e:4.9}). It can
be shown that, if we suppose that the matrices $P_i$ and $Q_i$ are of
maximum rank, then we get the trivial solution of the Toda equation
given by (\ref{e:4.2}). Hence, we will assume that $P_i$ and $Q_i$
are not of maximum rank. The simplest case here is given by matrices
of rank one which can be represented as
\begin{equation}
P_i = u^{}_i {}^{t\!} w^{}_i, \qquad Q_i = x^{}_i {}^{t\!} y^{}_i,
\label{**}
\end{equation}
where $u$, $w$, $x$ and $y$ are $n$-dimensional column vectors.

The $\bbZ$-gradation suggests that it is convenient to consider
the $n \times n$ matrix-valued functions $P_i$ and $Q_i$ in the
corresponding block-matrix form. According to the representation
(\ref{e:3.1}), we can write
\[
P_i = \left( \begin{array}{cccc}
(P_i)_{11} & (P_i)_{12} &\ldots& (P_i)_{1p} \\
(P_i)_{21} & (P_i)_{22} &\ldots& (P_i)_{2p} \\
\vdots & \vdots & \ddots & \vdots \\
(P_i)_{p1} & (P_i)_{p2} &\ldots& (P_i)_{pp}
\end{array} \right),
\]
and make similar block-matrix partition for $Q_i$, where the
submatrices $(P_i)_{\alpha\beta}$ and $(Q_i)_{\alpha\beta}$
are complex $n_\alpha \times n_\beta$ matrices. Then, in terms
of such block submatrices, the relations (\ref{**}) take the
forms
\[
(P_i)_{\alpha\beta} = u_{i,\alpha} \: {}^{t\!}w_{i,\beta},
\qquad
(Q_i)_{\alpha\beta} = x_{i,\alpha} \: {}^{t\!}y_{i,\beta},
\]
where the standard matrix multiplication of the $n_\alpha \times 1$
submatrices $u_{i,\alpha}$, $x_{i,\alpha}$ by the
$1 \times n_\beta$ submatrices ${}^{t\!}w_{i,\beta}$,
${}^{t\!}y_{i,\beta}$ is implied as respective. We see that, from the
point of view of the $\bbZ$-gradation, also the $n \times 1$ matrices
$u_i$, $w_i$, $x_i$ and $y_i$ receive a natural representation in a
block-matrix form,
\[
{}^{t\!}u_i = \left(
{}^{t\!}u_{i,1} \:\: \\
{}^{t\!}u_{i,2} \:\: \\
\ldots \\
{}^{t\!}u_{i,\alpha} \:\: \\
\ldots \\
{}^{t\!}u_{i,p}
\right),
\qquad
{}^{t\!}y_i = \left(
{}^{t\!}y_{i,1} \:\: \\
{}^{t\!}y_{i,2} \:\: \\
\ldots \\
{}^{t\!}y_{i,\alpha} \:\: \\
\ldots \\
{}^{t\!}y_{i,p}
\right),
\]
where $u_{i,\alpha}$ and $y_{i,\alpha}$, $\alpha=1,\ldots,p$,
are complex $n_\alpha \times 1$ matrices. We have similar
expressions also for $w_i$ and $x_i$. This representation,
together with the block-matrix form (\ref{*}) of $h$,
allows us to write the relations (\ref{e:4.8}) and
(\ref{e:4.9}) as follows:
\begin{gather}
{}^{t\!}y_{i,\alpha} + \sum_{j=1}^{r} \sum_{\delta,\beta=1}^p
\frac{\displaystyle
\nu_i \: \epsilon_p^{-\beta(\delta-\alpha)}}
{\displaystyle \nu_i - \epsilon_p^\beta \: \mu^{}_j}
\left( {}^{t\!}y_{i,\delta} \: u_{j,\delta} \right)
{}^{t\!}w_{j,\alpha} = 0,
\label{e:4.20} \\
u_{i,\alpha} + \sum_{j=1}^{r} \sum_{\delta,\beta=1}^p
\frac{\displaystyle
\mu_i \: \epsilon_p^{-\beta(\alpha-\delta)}}
{\displaystyle \mu_i - \epsilon_p^\beta \: \nu_j}
x_{j,\alpha} \left( {}^{t\!}y_{j,\delta} \: u_{i,\delta} \right) = 0.
\label{e:4.21}
\end{gather}
Using the identity
\begin{equation}
\sum_{\alpha=0}^{p-1} \frac{z \: \epsilon_p^{-\beta\alpha}}
{z - \epsilon_p^\alpha}
= p \frac{z^{p-|\beta|_p}}{z^p - 1},
\label{e:4.22}
\end{equation}
where $|\beta|_p$ is the residue of division of $\beta$ by $p$,
we can rewrite (\ref{e:4.20}) in terms of the block submatrices,
\begin{equation}
{}^{t\!}y_{i,\alpha} + p \sum_{j=1}^{r}
(R_\alpha)_{i j} \: {}^{t\!}w_{j,\alpha} = 0.
\label{e:4.23}
\end{equation}
Here the ${r} \times {r}$ matrices $R_\alpha$ are defined as
\[
(R_\alpha)_{i j} = \frac{1}{\nu_i^p - \mu_j^p}
\sum_{\beta=1}^p \nu_i^{p - |\beta-\alpha|_p} \mu_j^{|\beta-\alpha|_p}
\: {}^{t\!}y_{i,\beta} \: u_{j,\beta}.
\]
The identity (\ref{e:4.22}) allows us to write also the submatrix form
of (\ref{e:4.21}) as
\begin{equation}
u_{i,\alpha} + p \sum_{j=1}^{r} x_{j,\alpha} (S_\alpha)_{j i}
= 0, \label{e:4.24}
\end{equation}
where
\[
(S_\alpha)_{j i} = - \frac{1}{\nu^p_j - \mu^p_i}
\sum_{\beta=1}^p \nu_j^{|\alpha-\beta|_p}
\mu_i^{p - |\alpha-\beta|_p}
\: {}^{t\!}y_{j,\beta} \: u_{i,\beta}.
\]
With the help of the equality
\[
p - 1 - |\alpha-1|_p = |-\alpha|_p
\]
it is straightforward to demonstrate that
\[
(S_\alpha)_{j i} = - \frac{\mu_i}{\nu_j}
(R_{\alpha+1})_{j i},
\]
and so, (\ref{e:4.24}) can be written as
\begin{equation}
u_{i,\alpha} - p \mu_i \sum_{j=1}^{r} x_{j,\alpha}
\frac{1}{\nu_j} (R_{\alpha+1})_{j i} = 0.
\label{e:4.25}
\end{equation}
We use the equations (\ref{e:4.23}) and (\ref{e:4.25}) to express
the vectors $w_i$ and $x_i$ via the vectors $u_i$ and $y_i$,
\[
{}^{t\!}w_{i,\alpha} = - \frac{1}{p} \sum_{j=1}^{r}
(R^{-1}_\alpha)_{i j} \: {}^{t\!}y_{j,\alpha}, \qquad
x_{i,\alpha} = \frac{1}{p} \sum_{j=1}^{r}
u_{j,\alpha} \: \frac{1}{\mu_j} \:
(R^{-1}_{\alpha+1})_{j i} \: \nu_i.
\]
Apart from the summation over the pole indices $j$, there are the
corresponding matrix multiplications of the submatrices entering the
last two relations. As a result, we come to the following solution
of the relations (\ref{e:4.8}) and (\ref{e:4.9}):
\[
(P_i)_{\alpha\beta} = - \frac{1}{p} u_{i,\alpha} \sum_{j=1}^{r}
(R^{-1}_\beta)_{i j} \: {}^{t\!}y_{j,\beta}, \qquad
(Q_i)_{\alpha\beta} = \frac{1}{p} \sum_{j=1}^{r} u_{j,\alpha} \:
\frac{1}{\mu_j} \: (R^{-1}_{\alpha+1})_{j i} \: \nu_i \:
{}^{t\!}y_{i,\beta}.
\]
Using (\ref{e:4.6}) and (\ref{e:4.19}), we get
\[
\gamma = \psi^{}_\infty
= I_n + \sum_{i = 1}^{r} \sum_{\alpha=1}^p h^{\alpha} \: P_i \:
h^{-\alpha}.
\]
For the submatrices of $\gamma$ this gives the expression
\[
\gamma_{\alpha\beta}
= \delta_{\alpha\beta} \left( I_{n_\alpha} + p \sum_{i=1}^{r}
(P_i)_{\alpha \alpha} \right)
= \delta_{\alpha\beta} \left( I_{n_\alpha} - \sum_{i, j = 1}^{r}
u_{i,\alpha} \: (R^{-1}_\alpha)_{i j} \: {}^{t\!}y_{j,\alpha} \right).
\]
Hence, in view of the block-diagonal structure of $\gamma$, we have
\[
\Gamma_\alpha = 1 - \sum_{i, j = 1}^{r}
u_{i,\alpha} \: (R^{-1}_\alpha)_{i j} \: {}^{t\!}y_{j,\alpha}.
\]
According to our general convention, we assume that the
$n_\alpha \times 1$ matrix-valued functions $u_{i,\alpha}$ and
$y_{i,\alpha}$ are defined for arbitrary integer values of $\alpha$
and
\[
u_{i,\alpha+p} = u_{i,\alpha}, \qquad
y_{i,\alpha+p} = y_{i,\alpha}.
\]
The periodicity of $(R_\alpha)_{i j}$ actually follows from its
definition,
\[
(R_{\alpha+p})_{i j} = (R_{\alpha})_{i j}.
\]
It appears that it is more convenient to use quasi-periodic quantities
$\wt u_{i,\alpha}$, $\wt y_{i,\alpha}$ and $(\wt R_\alpha)_{i j}$
defined by
\begin{gather*}
\wt u_{i,\alpha} = u_{i,\alpha} \mu^\alpha_i, \qquad
\wt y_{i,\alpha} = y_{i,\alpha} \nu^{-\alpha}_i, \\
(\wt R_\alpha)_{i j} = \nu^{-\alpha}_i (R_\alpha)_{i j} \: \mu^\alpha_j.
\end{gather*}
For these quantities we have
\begin{gather*}
\wt u_{i,\alpha+p} = \wt u_{i,\alpha} \: \mu^p_i, \qquad
\wt y_{i,\alpha+p} = \wt y_{i,\alpha} \: \nu^{-p}_i, \\
(\wt R_{\alpha+p})_{i j} = \nu^{-p}_i
(\wt R_\alpha)_{i j} \: \mu^p_j.
\end{gather*}

The expression of the matrix elements of the matrices $\wt R_\alpha$
through the functions $\wt y_{i,\alpha}$ and $\wt u_{i,\alpha}$
has a nicely simplified form
\begin{equation}
(\wt R_\alpha)_{i j} = \frac{1}{\nu_i^p - \mu_j^p}
\left( \mu_j^p \sum_{\beta=1}^{\alpha-1}  {}^{t\!} \wt y_{i,\beta}
\: \wt u_{j,\beta}
+ \nu_i^p \sum_{\beta=\alpha}^p {}^{t\!} \wt y_{i,\beta}
\: \wt u_{j,\beta} \right).
\label{e:4.26}
\end{equation}
In terms of the quasi-periodic quantities, for the matrix-valued
functions $\Gamma_\alpha$ we have
\begin{equation}
\Gamma_\alpha = I_{n_\alpha} - \sum_{i,j=1}^{r}
\wt u_{i,\alpha} \: (\wt R^{-1}_\alpha)_{i j}
\: {}^{t\!} \wt y_{j,\alpha}.
\label{e:4.27}
\end{equation}
It is useful to have also the explicit expression of the inverse
mapping $\gamma^{-1}$. Using the relation
\[
\gamma^{-1} = \psi^{-1}_\infty
= I_n + \sum^{r}_{i=1} \sum^p_{\alpha=1}
h^{\alpha} \: Q_i \: h^{-\alpha}
\]
we derive
\begin{equation}
\Gamma^{-1}_\alpha = I_{n_\alpha} + \sum^{r}_{i,j=1}
\wt u^{}_{i,\alpha}
(\wt R^{-1}_{\alpha+1})_{i j} \: {}^{t\!} \wt y_{j,\alpha}.
\label{e:4.28}
\end{equation}
Using the definition of $\wt R_\alpha$, we come to the equality
\[
(\wt R^{}_{\alpha+1})_{i j} = (\wt R^{}_\alpha)_{i j}
- {}^{t\!} \wt y_{i,\alpha} \: \wt u_{j,\alpha}.
\]
It is clear that in the case under consideration we do not have
any determinant representation specific to the abelian case
\cite{Mik81, NirRaz08a}, and the last two relations are just
helpful for verifying the equations of motion by the obtained
solutions.

Further, it follows from (\ref{e:4.20}) and (\ref{e:4.21}) that,
to fulfill also (\ref{e:4.12})--(\ref{e:4.15}), it is sufficient
to satisfy the equations
\begin{gather}
\partial_- y_i = \nu_i^{-1} \: {}^{t\!}c_- \: y_i,
\qquad
\partial_+ y_i = \nu_i^{} \: {}^{t\!}c_+ \: y_i,
\label{e:4.29} \\
\partial_- u_i = - \mu_i^{-1} \: c_- \: u_i,
\qquad
\partial_+ u_i = - \mu_i^{} \: c_+ \: u_i.
\label{e:4.30}
\end{gather}
The general solution to these equations in the case when $c_-$ and
$c_+$ are constant is formally
\begin{gather*}
y_i(z^-,z^+)
= \exp(\nu_i^{-1} \: {}^{t\!}c_- \: z^-
+ \nu_i^{} \: {}^{t\!}c_+ \: z^+) \: y^0_i,
\\
u_i(z^-,z^+)
= \exp(- \mu_i^{-1} \: c_- \: z^-
- \mu_i^{} \: c_+ \: z^+) \: u^0_i,
\end{gather*}
where $y^0_i = y_i(0,0)$ and $u^0_i = u_i(0,0)$.

Thus we have shown that it is possible to satisfy (\ref{e:4.8}),
(\ref{e:4.9}) and (\ref{e:4.12})--(\ref{e:4.15}) and construct in
this way a wide class of solutions to the non-abelian loop Toda 
equations (\ref{e:3.4}). In what follows we will suppose that 
$c_-$ and $c_+$ represent constant mappings and shall make 
the above formal solution to the equations (\ref{e:4.29}), 
(\ref{e:4.30}) explicit.

\section{Deriving soliton solutions} \label{s:5}

\subsection{The eigenvalue problems} \label{s:5.1}

Seeing the formal expressions for $u_i$ and $y_i$, we understand
that we need to somehow handle the exponentials of the matrices
$c_-$ and $c_+$. To this end, it is customary to treat them as
matrices of linear operators. Assume that the submatrices entering
the mappings $c_-$ and $c_+$ are of maximum ranks, that is
\[
{\mathrm {rank}} \; C_{-\alpha} = {\mathrm{min}} \;
(n_{\alpha + 1},n_{\alpha}),
\qquad
{\mathrm {rank}} \; C_{+\alpha} = {\mathrm{min}} \;
(n_{\alpha},n_{\alpha + 1}),
\]
and they respect the commutativity of $c_-$ and $c_+$ according
to (\ref{e:4.1}). Here we consider the case where these matrices
are such that the corresponding $n_{\alpha+1} \times n_\alpha$ and
$n_{\alpha} \times n_{\alpha+1}$ submatrices $C_{-\alpha}$ and
$C_{+\alpha}$ can be brought to the forms
\[
\left( \begin{array}{c}
        I_{n_\alpha} \\ 0
       \end{array}
\right), \qquad
\left( I_{n_\alpha} \:\:\: 0 \right)
\]
if $n_\alpha \le n_{\alpha + 1}$, and
\[
\left( I_{n_{\alpha+1}} \:\:\: 0 \right), \qquad
\left( \begin{array}{c}
        I_{n_{\alpha+1}} \\ 0
       \end{array}
\right)
\]
if $n_{\alpha + 1} \le n_\alpha$, respectively, by implementing
the transformations (\ref{e:2.12}), or the same in the submatrix
form (\ref{e:3.6}), accompanied by an appropriate change of
independent variables.

Denote by $n_*$ the minimum value of the positive integers
$\{n_\alpha\}$. Consider the eigenvalue problems for the linear
operators $c_-$ and $c_+$. The corresponding characteristic
polynomial is $(-1)^n t^{n - p n_*} (t^p - 1)^{n_*}$ giving
rise to the characteristic equation
\[
t^{n-pn_*} \prod^{p}_{\alpha=1}
\left( t - \epsilon_p^\alpha \right)^{n_*} = 0.
\]
Therefore, the spectrum consists of the zero eigenvalue
having the algebraic multiplicity $n-pn_*$ and nonzero
eigenvalues being powers of the $p$th root of unity
having the algebraic multiplicity $n_*$ each. We also
take into account that the spectra of similar matrices
coincide.

The eigenvalue problem relations
\[
c_- \Psi_{\beta} = \epsilon_p^{-\beta} \Psi_{\beta}, \qquad
c_+ \Psi_{\beta} = \epsilon_p^\beta \Psi_{\beta}
\]
are satisfied by the eigenvectors\footnote{Here $\Psi_\beta$ is in
fact an $n \times n_*$ matrix being thus a collection of different
$n_*$ eigenvectors of $c_-$ corresponding to one and the same
eigenvalue $\epsilon^{-\beta}_p$.}
\[
{}^{t\!}\Psi_{\beta} = \left(
{}^{t\!}\Psi_{\beta,1},\ldots,
{}^{t\!}\Psi_{\beta,\alpha},\ldots,
{}^{t\!}\Psi_{\beta,p}
\right)
\]
with
\[
\Psi_{\beta,\alpha} = \epsilon_p^{\alpha \beta} \theta_\alpha,
\]
where constant $n_\alpha \times n_*$ submatrices $\theta_\alpha$
are subject to the conditions
\begin{equation}
C_{-\alpha} \theta_{\alpha} = \theta_{\alpha+1}, \qquad
C_{+\alpha} \theta_{\alpha+1} = \theta_\alpha.
\label{e:5.1}
\end{equation}
Denote by $k$ the rank of the matrix $c_-$.
Then we have $n-k={\mathrm{dim}\;{\mathrm{ker}\;c_-}}$.
It is clear that for the case under consideration
$k = \mathrm{rank}\;c_-
= \sum^p_{\alpha=1}\mathrm{min}\;(n_\alpha,n_{\alpha+1}) \ge pn_*$.
In general, the algebraic multiplicity of an eigenvalue does not
coincide with its geometric multiplicity, the former is just
non less than the latter. To be precise, here we have that the
algebraic and geometric multiplicities of the nonzero eigenvalues
$t = \epsilon_p^{\pm \alpha}$ are one and the same and equal
to $n_*$ for all $\alpha=1,2,\ldots,p$. Indeed, it can be shown
that there are no generalized eigenvectors of $c_-$ corresponding
to its nonzero eigenvalues, that is, no nontrivial eigenvectors of
the form $\Psi'_\beta = (c_- - \epsilon^{-\beta}_p I_n) \Psi_\beta$
for any $\beta = 1,2,\ldots,p$. In contrast, there are nontrivial
generalized eigenvectors of $c_-$ corresponding to the zero eigenvalue.
As a consequence, the algebraic multiplicity of the zero eigenvalue,
equal to $n - pn_*$ as is seen from the characteristic equation,
does not coincide with its geometric multiplicity equal to $n - k$.

Hence, treating $c_-$ as a matrix of a linear operator acting on an
$n$-dimensional vector space $V$, we see that the latter can be
decomposed into a direct sum as
\[
V = V_0 \oplus V_1,
\]
where $V_1$ is a $pn_*$-dimensional subspace spanned by
the $\Psi$-eigenvectors of $c_-$ with nonzero eigenvalues,
actually, $V_1 = {\mathrm{im}\;c_-}$, and $V_0$ is simply
defined to be its orthogonal complement spanned by the null
vectors and generalized null vectors of $c_-$. In this, the
null-subspace spanned by the generalized null vectors has
the dimension $k - p n_*$.

Similar consideration can be given for the matrix $c_+$ as well.
Note also that the above decomposition induces the corresponding
dual decomposition.

Now, seeing the structure of the general solution for $y_i$ and
$u_i$, we can proceed as follows. Expand the initial values
$y^0_i$ and $u^0_i$ over the basis vectors of $V$, taking
into account its decomposition:
\[
y^0_i = \Psi_{0} \: d_{i,0}
+ \sum^p_{\alpha=1} \Psi_{\alpha} \: d_{i,\alpha},
\qquad
u^0_i = \Psi_{0} \: c_{i,0}
+ \sum^p_{\alpha=1} \Psi_{\alpha} \: c_{i,\alpha},
\]
where $\Psi_{0}$ is an $n \times (n-pn_*)$ matrix whose columns are
an appropriate collection of basis vectors of $V_0$, the $n \times n_*$
matrices $\Psi_{\alpha}$ are giving basis vectors of $V_1$ as introduced
earlier, $c_{i,0}$ and $d_{i,0}$ are $(n-pn_*) \times 1$
matrices, while $c_{i,\alpha}$ and $d_{i,\alpha}$ are $n_* \times 1$
matrices, altogether encoding the initial value data for $u_i$ and
$y_i$. Then, in view of the above consideration of the properties of
the matrices $c_\pm$, the general solutions to (\ref{e:4.29}),
(\ref{e:4.30}) take the forms
\begin{gather}
y_i(z^-,z^+) = \Psi_{0} \: q_{i-}(z^-) q_{i+}(z^+) \: d_{i,0}
+ \sum^p_{\alpha=1} \Psi_{\alpha}
\exp \left( \nu^{-1}_i \: \epsilon_p^{\alpha} \: z^-
+ \nu_i \: \epsilon_p^{-\alpha} \: z^+ \right) d_{i,\alpha},
\label{e:5.2} \\
u_i(z^-,z^+) = \Psi_{0} \: p_{i-}(z^-) p_{i+}(z^+) \: c_{i,0}
+ \sum^p_{\alpha=1} \Psi_{\alpha}
\exp \left( -\mu^{-1}_i \: \epsilon_p^{-\alpha} \: z^-
- \mu_i \: \epsilon_p^\alpha \: z^+ \right)
c_{i,\alpha},
\label{e:5.3}
\end{gather}
where $p_{i\pm}(z^\pm)$ and $q_{i\pm}(z^\pm)$ are
$(n-pn_*) \times (n-pn_*)$ matrices with matrix elements
$\left(p_{i\pm}(z^\pm)\right)_{ab}$ and
$\left(q_{i\pm}(z^\pm)\right)_{ab}$,
$a,b=1,2,\ldots,n-pn_*$, being polynomials in $z^\pm$ of
degrees not greater than $k - p n_*$. More specifically, these
polynomials are such that the equations (\ref{e:4.29}) and
(\ref{e:4.30}) should be satisfied.

Further, for the submatrices introduced earlier according
to the $\bbZ$-grading structure we obtain
\begin{multline*}
{}^{t\!}y_{i,\delta} \: u_{j,\delta}
=  {}^{t\!}d_{i,0} \: {}^{t\!}q_{i+} {}^{t\!}q_{i-}
({}^{t\!}\Psi_{0,\delta} \Psi_{0,\delta}) p_{j-} p_{j+}
\: c_{j,0} \\[.5em]
 + \sum^p_{\alpha,\beta=1}
\exp{\left( Z_{-\alpha}(\nu_i) -  Z_{\beta}(\mu_j) \right)}
{}^{t\!}d_{i,\alpha} \: ({}^{t\!}\Psi_{\alpha,\delta}
\: \Psi_{\beta,\delta}) \: c_{j,\beta},
\end{multline*}
where for convenience we have introduced the notation
\[
Z_\alpha(\mu_i) = \mu^{-1}_i \: \epsilon_p^{-\alpha} \: z^-
+ \mu^{}_i \: \epsilon_p^\alpha \: z^+.
\]
Recalling that
$\Psi_{\beta,\delta} = \epsilon_p^{\delta\beta} \theta_\delta$
and passing to quasi-periodic quantities, we find
\begin{multline*}
{}^{t\!} \wt y_{i,\delta} \: \wt u_{j,\delta}
=  {}^{t\!}d_{i,0} \: {}^{t\!}q_{i+} {}^{t\!}q_{i-}
\nu^{-\delta}_i \: ({}^{t\!}\Psi_{0,\delta} \Psi_{0,\delta})
\: \mu^\delta_j \: p_{j-} p_{j+} \: c_{j,0} \\[.5em]
+ \sum^p_{\alpha,\beta=1} \nu^{-\delta}_i \: \mu^\delta_j
\: \epsilon_p^{\delta(\alpha + \beta)}
\exp{\left( Z_{-\alpha}(\nu_i) - Z_{\beta}(\mu_j) \right)}
{}^{t\!}d_{i,\alpha} \: ({}^{t\!}\theta_\delta \theta_\delta) \:
c_{j,\beta}.
\end{multline*}
Suppose that the $n_\delta \times n_*$ submatrices $\theta_\delta$ are
such that
\[
{}^{t\!}\theta_\delta \theta_\delta = \Theta,
\]
where $\Theta$ is one and the same non-degenerate $n_* \times n_*$ matrix
for all $\delta=1,2,\ldots,p$. Consequently, we have from (\ref{e:4.26})
\begin{eqnarray}
(\wt R_\alpha)_{i j} &=&
{}^{t\!}d_{i,0} \: {}^{t\!}q_{i+} {}^{t\!}q_{i-}
(\Psi^0_\alpha)^{}_{i j} \: p_{j-} p_{j+} \: c_{j,0}
\nonumber \\[.5em]
&& + \sum^p_{\beta,\delta=1} \rme^{Z_{-\beta}(\nu_i)
- Z_\delta(\mu_j)}
\frac{\epsilon_p^{\alpha(\beta+\delta)}}{1-\mu_j\nu^{-1}_i
\epsilon_p^{\beta+\delta}} \:
{}^{t\!}d_{i,\beta} \: \Theta \: c_{j,\delta} \:
(\mu^\alpha_j \: \nu^{-\alpha}_i),
\label{e:5.4}
\end{eqnarray}
where, for sake of brevity, we have used the notation
\[
(\Psi^0_\alpha)^{}_{i j} = \frac{1}{\nu^p_i - \mu^p_j}
\left( \mu^p_j \sum^{\alpha-1}_{\varepsilon=1}
\nu^{-\varepsilon}_i \:
{}^{t\!}\Psi_{0,\varepsilon} \: \Psi_{0,\varepsilon}
\: \mu^\varepsilon_j + \nu^p_i
\sum^p_{\varepsilon=\alpha} \nu^{-\varepsilon}_i \:
{}^{t\!}\Psi_{0,\varepsilon} \: \Psi_{0,\varepsilon}
\: \mu^\varepsilon_j \right)
\]
for this constant $(n-pk_*) \times (n-pk_*)$ matrix.
The relations (\ref{e:5.2}), (\ref{e:5.3}) and (\ref{e:5.4})
allow us to construct general solutions $\Gamma_\alpha$ by
(\ref{e:4.27}) (it should be instructive to compare our
general solution with the corresponding construction of
\cite{EtiGelRet97a} where the notion of quasi-determinants
was exploited for the purpose).

\subsection{One-soliton solution} \label{s:5.2}

To construct simplest one-soliton solutions ($r = 1$) to the Toda
equations, we assume that the initial data of the system are such
that the coefficients $c_{\alpha}$ are nonzero only for one value of
the index $\alpha$, which we denote by $I$, and the coefficients
$d_{\alpha}$ are nonzero only for two values of the index $\alpha$,
which we denote by $J$ and $K$. Besides, let $c_{0}$ and $d_{0}$
be zero.\footnote{Note that, essentially unlike the twisted abelian
case \cite{NirRaz08b}, keeping these null-eigenspace coefficients
nonzero here we do not obtain any soliton-like solutions.} Thus,
for $n_\alpha \times 1$ submatrices of the matrix-valued functions
$\wt u$ and $\wt y$ we have
\begin{eqnarray*}
&& \wt u^{}_{\alpha} = \mu^{\alpha} \: \epsilon_p^{\alpha I}
\: \rme^{-Z_I(\mu)} \: \theta_\alpha \: c_{I},
\\[.5em]
&& {}^{t\!}\wt y_{\alpha} = \nu^{-\alpha} \: \epsilon_p^{\alpha J}
\: \rme^{Z_{-J}(\nu)} \: {}^{t\!}d_{J} \: {}^{t\!}\theta_\alpha
+ \nu^{-\alpha} \: \epsilon_p^{\alpha K} \: \rme^{Z_{-K}(\nu)}
\: {}^{t\!}d_{K} \: {}^{t\!}\theta_\alpha.
\end{eqnarray*}
The matrix $\wt R_\alpha$ is simply a function for this case and
is defined by the expression
\begin{multline*}
\wt R_\alpha = \mu^\alpha \: \nu^{-\alpha} \: \epsilon_p^{I\alpha}
\: \rme^{-Z_I(\mu)} \left(
\rme^{Z_{-J}(\nu)} \frac{\epsilon_p^{J\alpha}}
{1 - \mu \nu^{-1} \epsilon_p^{I+J}}
\: ({}^{t\!}d_{J} \: \Theta \: c_{I}) \right. \\
\left. + \rme^{Z_{-K}(\nu)} \frac{\epsilon_p^{K\alpha}}
{1 - \mu \nu^{-1} \epsilon_p^{I+K}}
\: ({}^{t\!}d_{K} \: \Theta \: c_{I})
\right).
\end{multline*}
And for the $n_\alpha \times n_\alpha$ matrix-valued functions
$\Gamma_\alpha$ this gives
\[
\Gamma_\alpha = \frac{I_{n_\alpha}
- (1 - \mu \nu^{-1} \epsilon_p^{I+J}) Y^{(J)}_\alpha
+ \wt d \epsilon_p^{\alpha(K-J)}
\rme^{Z_{-K}(\nu) - Z_{-J}(\nu)}(I_{n_\alpha}
- (1 - \mu \nu^{-1} \epsilon_p^{I+K}) Y^{(K)}_\alpha)}
{1 + \wt d \epsilon_p^{\alpha(K-J)} \rme^{Z_{-K}(\nu) - Z_{-J}(\nu)}},
\]
where we have introduced constant idempotent
$n_\alpha \times n_\alpha$ matrices
\[
Y^{(A)}_\alpha = \frac{(\theta_\alpha \: c_{I} \: {}^{t\!}d_{A}
\: {}^{t\!}\theta_\alpha)}{{}^{t\!}d_{A} \: \Theta \: c_{I}},
\qquad
(Y^{(A)}_\alpha)^2 = Y^{(A)}_\alpha, \qquad A = J, K,
\]
satisfying the relations
\[
Y^{(J)}_\alpha Y^{(K)}_\alpha = Y^{(K)}_\alpha, \qquad
Y^{(K)}_\alpha Y^{(J)}_\alpha = Y^{(J)}_\alpha,
\]
and also the notation
\[
\wt d = \frac{{}^{t\!}d_{K} \: \Theta \: c_{I}}
{{}^{t}d_{J} \: \Theta \: c_{I}}
\frac{1 - \mu \nu^{-1} \epsilon_p^{I+J}}
{1 - \mu \nu^{-1} \epsilon_p^{I+K}}.
\]
The expression for $\Gamma_\alpha$ can be rewritten as
\[
\Gamma_\alpha = \left[ I_{n_\alpha} -
(1 - \mu \nu^{-1} \epsilon_p^{I+J})Y^{(J)}_\alpha \right]
\frac{I_{n_\alpha} + \epsilon_p^{\alpha\rho}
\rme^{Z(\zeta) + \wt\delta} \wt X_\alpha}
{1 + \epsilon_p^{\alpha\rho} \rme^{Z(\zeta) + \wt\delta}},
\]
where
\[
\wt X_\alpha
= \left( I_{n_\alpha} - (1 - \mu \nu^{-1} \epsilon_p^{I+J})
Y^{(J)}_\alpha \right)^{-1}
\left( I_{n_\alpha} - (1 - \mu \nu^{-1} \epsilon_p^{I+K})
Y^{(K)}_\alpha \right)
\]
and we use the notations
\[
\rho = K-J, \qquad \kappa_\rho = 2\sin\frac{\pi\rho}{p}, \qquad
\zeta = -\rmi \nu \epsilon_p^{-(K+J)/2}, \qquad \wt d = \exp\wt\delta,
\]
and the function $Z$ in the exponent takes the most familiar form
\[
Z(\zeta) = \kappa_{\rho} ( \zeta^{-1} \: z^- + \zeta \: z^+ ).
\]
With the help of the above properties of the matrices
$Y^{(J)}_\alpha$, it is not difficult to make sure that
\[
h^{}_{\alpha,J} C_{+\alpha} h^{-1}_{\alpha+1,J} = C_{+\alpha}
\]
for
\[
h_{\alpha,J} = I_{n_\alpha} - (1 - \mu \nu^{-1} \epsilon_p^{I+J})
Y^{(J)}_\alpha, \qquad
h^{-1}_{\alpha,J} = I_{n_\alpha}
- (1 - \mu^{-1} \nu \epsilon_p^{-(I+J)}) Y^{(J)}_\alpha.
\]
Therefore the transformation
\begin{equation}
h_{\alpha,J} \Gamma_\alpha \to \Gamma_\alpha
\label{***}
\end{equation}
is a symmetry transformation of the Toda equations (\ref{e:3.5})
realizing a particular case of the simplest WZNW-type symmetry
transformation described by the relations (\ref{e:3.6}) with
$\eta_{+ \alpha} = h_{\alpha,J}$, $\eta_{- \alpha} = I_{n_\alpha}$.
Similarly, one can use the relations
\[
h^{-1}_{\alpha+1,J} C_{-\alpha} h_{\alpha,J} = C_{-\alpha}
\]
to show that also the transformation
\[
\Gamma_\alpha h_{\alpha,J} \to \Gamma_\alpha
\]
is a symmetry transformation of the Toda equations (\ref{e:3.5})
corresponding to the transformations (\ref{e:3.6}) with
$\eta_{- \alpha} = h^{-1}_{\alpha,J}$, $\eta_{+ \alpha} = I_{n_\alpha}$.

Now, performing the symmetry transformation (\ref{***}), we write
the one-soliton solution to the equations (\ref{e:3.5}) as follows:
\[
\Gamma_\alpha = \frac{I_{n_\alpha} + \epsilon_p^{\alpha\rho}
\rme^{Z(\zeta) + \wt\delta} \wt X_\alpha}
{1 + \epsilon_p^{\alpha\rho} \rme^{Z(\zeta) + \wt\delta}},
\]
with all entries defined above. Using the properties of the
idempotent matrices $Y^{(A)}_\alpha$, we can also rewrite the
expressions for the matrices $\wt X_\alpha$ as
\[
\wt X_\alpha = h^{-1}_{\alpha,J} h_{\alpha,K}
= I_{n_\alpha} + \mu^{-1} \nu \epsilon_p^{-(I+J)}
\left( (1-\mu \nu^{-1} \epsilon_p^{I+J} ) Y^{(J)}_\alpha
- (1-\mu \nu^{-1} \epsilon_p^{I+K}) Y^{(K)}_\alpha \right).
\]
Using (\ref{e:4.28}) we can also show that
\[
\Gamma_\alpha^{-1} = \frac{I_{n_\alpha} + \epsilon_p^{(\alpha+1)\rho}
\rme^{Z(\zeta) + \wt\delta} \wt X^{\prime}_\alpha}
{1 + \epsilon_p^{(\alpha+1)\rho} \rme^{Z(\zeta) + \wt\delta}},
\]
where
\[
\wt X^{\prime}_{\alpha} = h^{-1}_{\alpha,K} \: h^{}_{\alpha,J}.
\]
Here it is obvious that
$\wt X^{\prime}_{\alpha} = \wt X^{-1}_{\alpha}$,
and it is not difficult to show that
$\Gamma^{-1}_\alpha \Gamma^{}_\alpha = I_{n_\alpha}$.
It is worthwhile noting that when $n_\alpha = 1$, we get
$Y^{(J)}_\alpha = Y^{(K)}_\alpha = 1$, so that
$\wt X_\alpha = \epsilon_p^\rho$, while
$\wt X'_\alpha = \epsilon_p^{-\rho}$, and also $\wt d \to d$,
and so we recover precisely the abelian case \cite{NirRaz08a}.

\subsection{Multi-soliton solutions} \label{s:5.3}

Now, to obtain solutions depending on $r$ linear combinations of
independent variables we assume that for each value of the index
$i = 1,\ldots,r$ the matrix-valued coefficients $c_{i,\alpha}$ are
different from zero for only one value of $\alpha$, which we denote
by $I_i$, and that the matrix-valued coefficients $d_{i,\alpha}$ are
different from zero for only two values of $\alpha$, which we denote
by $J_i$ and $K_i$. And we also use the following slightly simplified
notation for such nonvanishing initial-data $n_* \times 1$
matrix-valued coefficients:
\[
d_{J_i} = d_{i,J_i}, \quad d_{K_i} = d_{i,K_i} \qquad
c_{I_i} = c_{i,I_i}.
\]

Then we have from the equality (\ref{e:5.4}) that
\begin{multline*}
(\wt R_\alpha)_{i j} =
\nu^{-\alpha}_i \: \epsilon_p^{\alpha J_i} \: \rme^{Z_{-J_i}(\nu_i)}
\left( \frac{{}^{t\!}d_{J_i} \: \Theta \: c^{}_{I_j}}
{1 - \mu_j \nu^{-1}_i \epsilon_p^{I_j + J_i}}
\right. \\[.5em]
 \left. + \epsilon_p^{(K_i - J_i)\alpha} \:
\rme^{Z_{-K_i}(\nu_i) - Z_{-J_i}(\nu_i)}
\frac{{}^{t\!}d_{K_i} \: \Theta \: c^{}_{I_j}}
{1-\mu_j \nu^{-1}_i \epsilon_p^{I_j + K_i}} \right)
\mu^\alpha_j \: \epsilon_p^{\alpha I_j} \: \rme^{-Z_{I_j}(\mu_j)}.
\end{multline*}
With account of explicit forms of $u^{}_{i,\alpha}$ and
${}^{t\!}y_{j,\alpha}$, the expression for $\Gamma_\alpha$
can be written in the form
\[
\Gamma_\alpha = I_{n_\alpha} - \sum^{r}_{i,j=1}
(\wt R^{\prime -1}_\alpha)^{}_{i j}
\left( \wt Y^{(J)}_{\alpha,j i}
+ \epsilon_p^{\alpha\rho_j} \: \rme^{Z(\zeta_j)}
\: \wt Y^{(K)}_{\alpha,j i} \right),
\]
where
\[
(\wt R'_\alpha)^{}_{i j}
= \wt D_{i j}(\nu \epsilon_p^{-J},\mu \epsilon_p^I)
+ \epsilon_p^{\alpha\rho_i} \: \rme^{Z(\zeta_i)}
\: \wt D_{i j}(\nu \epsilon_p^{-K},\mu \epsilon_p^I)
\]
and (for $A = J, K$)
\[
\wt D_{i j}(\nu \epsilon_p^{-A},\mu \epsilon_p^I)
= ({}^{t\!}d_{A_i} \: \Theta \: c^{}_{I_j}) \:
D_{i j}(\nu \epsilon_p^{-A},\mu \epsilon_p^I)
= ({}^{t\!}d_{A_i} \: \Theta \: c^{}_{I_j})
\frac{\nu_i \: \epsilon_p^{-A_i}}
{\nu_i \: \epsilon_p^{-A_i} - \mu_j \: \epsilon_p^{I_j}},
\]
(cp. with the notation used for the abelian case \cite{NirRaz08a})
and besides,
\[
\wt Y^{(A)}_{\alpha,i j} = ({}^{t\!}d_{A_i} \: \Theta \: c^{}_{I_j})
Y^{(A)}_{\alpha,i j} = \theta_\alpha \: c^{}_{I_j}
\: {}^{t\!}d_{A_i} \: {}^{t\!}\theta_\alpha.
\]
The idempotent $n_\alpha \times n_\alpha$ matrices
$Y^{(A)}_{\alpha,i j}$ satisfy the following remarkable
properties:
\begin{equation}
Y^{(A)}_{\alpha,i j} Y^{(B)}_{\alpha,k\ell} =
\frac{{}^{t\!}d_{A_i} \Theta c_{I_\ell}}
{{}^{t\!}d_{A_i} \Theta c_{I_j}} \cdot
\frac{{}^{t\!}d_{B_k} \Theta c_{I_j}}
{{}^{t\!}d_{B_k} \Theta c_{I_\ell}}
Y^{(B)}_{\alpha,k j}, \qquad A,B = J,K,
\label{e:5.5}
\end{equation}
while their tilded counterparts are subject to the relations of
simpler forms,
\begin{equation}
\wt Y^{(A)}_{\alpha,i j} \: \wt Y^{(B)}_{\alpha,k\ell} =
({}^{t\!}d_{A_i} \: \Theta \: c_{I_\ell}) \wt Y^{(B)}_{\alpha,k j}.
\label{e:5.6}
\end{equation}

Further, we can write
\[
\Gamma_\alpha = \frac{\displaystyle
I_{n_\alpha} \det \wt R'_\alpha - \sum^{r}_{i,j=1}
         (\wt\calR'_\alpha)^{}_{i j}
\left( \wt Y^{(J)}_{\alpha,i j} + \epsilon_p^{\alpha\rho_i}
\: \rme^{Z(\zeta_i)} \: \wt Y^{(K)}_{\alpha,i j}\right)}
{\displaystyle \det \wt R'_\alpha},
\]
meaning that, according to Leibniz,
\[
\det \wt R'_\alpha = \sum_{\sigma \in S_{r}} {\mathrm{sgn}}(\sigma)
\prod^{r}_{\ell=1} \left(
\wt D^{}_{\ell,\sigma(\ell)}(\nu \epsilon_p^{-J},\mu \epsilon_p^I)
+ \epsilon_p^{\alpha\rho_\ell} \: \rme^{Z(\zeta_\ell)} \:
\wt D^{}_{\ell,\sigma(\ell)}(\nu \epsilon_p^{-K},\mu \epsilon_p^I)
\right)
\]
and
\[
(\wt\calR'_\alpha)^{}_{i j}
= \sum_{\scriptstyle \sigma \in S_{r}}
{\mathrm{sgn}}(\sigma)
\prod^{r}_{\scriptstyle \ell=1 \atop
\scriptstyle \ell \ne i, \sigma(\ell) \ne j}
\left( \wt D^{}_{\ell,\sigma(\ell)}
(\nu \epsilon_p^{-J},\mu \epsilon_p^I)
+ \epsilon_p^{\alpha\rho_\ell} \: \rme^{Z(\zeta_\ell)} \:
\wt D^{}_{\ell,\sigma(\ell)}(\nu \epsilon_p^{-K},\mu \epsilon_p^I)
\right).
\]
Here $S_{r}$ is the symmetric group on the set of integers
$\{1,2,\ldots,{r}\}$, and $\mathrm{sgn}(\sigma)$ denotes
the signature of the permutation $\sigma$.

For sake of brevity, it is also convenient to denote
$\wt D_{i j}(A) = \wt D_{i j}(\nu \epsilon_p^{-A},\mu \epsilon_p^{I})$.
Seeing that
\[
\det \wt R'_\alpha =
\det \wt D(\nu \epsilon_p^{-J},\mu \epsilon_p^{I}) \cdot
\det \wt R''_\alpha,
\]
where
\[
(\wt R''_\alpha)^{}_{i j} = \delta_{i j}
+ \epsilon_p^{\alpha\rho_i} \: \rme^{Z(\zeta_i)} \sum^{r}_{k=1}
\wt D^{}_{i k}(\nu \epsilon_p^{-K},\mu \epsilon_p^{I})
\wt D^{-1}_{k j}(\nu \epsilon_p^{-J},\mu \epsilon_p^{I}),
\]
we can also write
\[
\Gamma_\alpha = \frac{\displaystyle
I_{n_\alpha} \det \wt R''_\alpha - \sum^{r}_{i,j,k=1}
    \wt D^{-1}_{i k}(J) (\wt\calR''_\alpha)^{}_{k j}
\left( \wt Y^{(J)}_{\alpha,j i} + \epsilon_p^{\alpha\rho_j}
\: \rme^{Z(\zeta_j)} \: \wt Y^{(K)}_{\alpha,j i}\right)}
{\displaystyle \det \wt R''_\alpha},
\]
where
\[
(\wt\calR''_\alpha)^{}_{k j}
= \sum_{\scriptstyle \sigma \in S_{r}}
{\mathrm{sgn}}(\sigma)
\prod^{r}_{\scriptstyle \ell = 1 \atop
\scriptstyle \ell \ne j, \sigma(\ell) \ne k}
\left( \delta_{\ell,\sigma(\ell)}
+ \epsilon_p^{\alpha\rho_\ell} \: \rme^{Z(\zeta_\ell)}
\sum^{r}_{m=1} \wt D^{}_{\ell m}(K)
\wt D^{-1}_{m,\sigma(\ell)}(J)
\right).
\]
Let us introduce a new notation for sake of certain brevity:
\[
\wt H^{}_{i j}(K,J) = \sum^{r}_{k=1}
\wt D^{}_{i k}(K) \: \wt D^{-1}_{k j}(J).
\]
Then we find that
\[
\Gamma_\alpha = h_{\alpha,J} \: {\wt T^{-1}_\alpha} \: {\wt T^X_\alpha},
\]
where
\[
h_{\alpha,J} = I_{n_\alpha} - \sum^{r}_{i,j=1}
\wt D^{-1}_{i j}(J) \: \wt Y^{(J)}_{\alpha,j i},
\]
and the quantities $\wt T_\alpha$ and $n_\alpha \times n_\alpha$
matrices $\wt T^X_\alpha$ together represent a non-abelian analogue
of the Hirota's $\tau$-functions,
\begin{gather}
 \wt T_\alpha = 1 + \sum^{r}_{i=1} E_{\alpha,i}
+ \sum^{r}_{\ell=2}
\sum^{r}_{\scriptstyle i_1,i_2,\ldots,i_\ell=1
\atop \scriptstyle i_1 < i_2 < \ldots < i_\ell}
\wt \eta_{i_1 i_2\ldots i_\ell} \:
E_{\alpha,i_1} \: E_{\alpha,i_2} \ldots E_{\alpha,i_\ell},
\nonumber \\
 \label{e:5.7} \\
 \wt T^X_\alpha = I_{n_\alpha} + \sum^{r}_{i=1} E_{\alpha,i} \:
\wt X_{\alpha,i}
+ \sum^{r}_{\ell=2}
\sum^{r}_{\scriptstyle i_1,i_2,\ldots,i_\ell=1
\atop \scriptstyle i_1 < i_2 < \ldots < i_\ell}
\wt \eta_{i_1 i_2 \ldots i_\ell} \:
E_{\alpha,i_1} \: E_{\alpha,i_2} \ldots E_{\alpha,i_\ell}
\wt X_{\alpha, i_1 i_2 \ldots i_\ell}.
\nonumber
\end{gather}
Here we also use our standard notation, coming yet from the abelian
case \cite{NirRaz08a},
\[
E_{\alpha,i} = \epsilon_p^{\alpha\rho_i} \rme^{Z(\zeta_i)
+ \wt \delta_i},
\]
with the quantities $\wt \delta_i$ defined by
\[
\rme^{\wt \delta_i} \equiv \wt H^{}_{i i}
= \sum^{r}_{k=1}
\wt D^{}_{i k}(K) \: \wt D^{-1}_{k i}(J),
\]
and the `soliton interaction coefficients' given by
\[
\wt \eta_{i_1 i_2 \ldots i_\ell} =
\frac{\displaystyle \sum_{\sigma \in S_\ell} \mathrm{sgn}(\sigma)
\prod^\ell_{m=1} \wt H_{i_m i_{\sigma(m)}}}
{\displaystyle \prod^\ell_{k=1} \wt H_{i_k i_k}}.
\]
Note that in the abelian case \cite{NirRaz08a} these quantities
factorize into pairwise interaction coefficients. To be noted
also here the relationship with our former notation from
\cite{NirRaz08a}, $\wt d_i \equiv \wt H_{i i}$. Finally,
the matrices $\wt X_{\alpha, \cdots }$ are defined as
follows:
\[
\wt X_{\alpha,i_1 i_2 \ldots i_\ell}
= h^{-1}_{\alpha,J} \: X_{\alpha,i_1 i_2 \ldots i_\ell},
\]
where for $\ell = 1$
\[
X_{\alpha,i} = I_{n_\alpha} - \frac{1}{\wt H^{}_{i i}}
\sum^{r}_{j,k=1} \left(
\wt D^{-1}_{k j}(J) \: \wt H^{}_{i i}
- \wt D^{-1}_{k i}(J) \: \wt H^{}_{i j}
\right) \wt Y^{(J)}_{\alpha,j k}
- \frac{1}{\wt H^{}_{i i}}
\sum^{r}_{k=1} \wt D^{-1}_{k i}(J)
\: \wt Y^{(K)}_{\alpha,i k}
\]
and the higher order quantities are
\begin{eqnarray*}
X_{\alpha,i_1 i_2 \ldots i_\ell} &=& I_{n_\alpha}
- \frac{1}{\displaystyle \sum_{\sigma \in S_\ell} \mathrm{sgn}(\sigma)
\displaystyle \prod^\ell_{m=1} \wt H^{}_{i_m i_{\sigma(m)}}}
\times \\
&& \times \left( \sum^{r}_{j,k=1}
\sum_{\sigma \in S'_{\ell+1}}
\mathrm{sgn}(\sigma) \:
\wt D^{-1}_{k \sigma(j)}(J) \:
\wt H^{}_{i_1 i_{\sigma(1)}}
\wt H^{}_{i_2 i_{\sigma(2)}} \ldots
\wt H^{}_{i_\ell i_{\sigma(\ell)}} \:
\wt Y^{(J)}_{\alpha,j k} \right. \\
&& \hskip5mm
+ \left. \sum^{r}_{k=1} \sum_{\sigma \in S_\ell}
\mathrm{sgn}(\sigma) \:
\wt D^{-1}_{k i_{\sigma(1)}}(J) \:
\wt H^{}_{i_2 i_{\sigma(2)}}
\wt H^{}_{i_3 i_{\sigma(3)}} \ldots
\wt H^{}_{i_\ell i_{\sigma(\ell)}} \:
\wt Y^{(K)}_{\alpha, i_1 k} \right), \quad \ell \ge 2,
\end{eqnarray*}
with $S_\ell$ being the symmetric group on the set
$\{1,2,\ldots,\ell\}$, $S'_{\ell + 1}$ the symmetric
group on the set
$\{j, i_1, i_2, \ldots, i_\ell\}$,
everywhere admitting that
$\sigma(i_m) = i_{\sigma(m)}$.

In particular, for $\ell=r$, we obtain a remarkably
simple equality
\[
X_{\alpha,1 2 \ldots {r}} = h_{\alpha,K}
= I_{n_\alpha} - \sum^{r}_{j,k=1}
\wt D^{-1}_{k j}(K) \:
\wt Y^{(K)}_{\alpha,j k}
\]
and
\[
\wt \eta^{}_{12\ldots{r}}
= \frac{\displaystyle \det \wt H}{\displaystyle \prod^{r}_{k=1}
\wt H_{i_k i_k}}
= \exp{\biggl( -\sum^{r}_{k=1} \wt \delta_k \biggr)}
\: \det \wt H.
\]
We notice that $h^{-1}_{\alpha,J}$ is a linear combination of
$\wt Y^{(J)}_{\alpha,i j}$,
\[
h^{-1}_{\alpha,J} = I_{n_\alpha} + \sum^{r}_{i,j=1}
\wt B^{-1}_{j i}(J) \: \wt Y^{(J)}_{\alpha,i j},
\qquad
\wt B_{i j}(J) = ({}^{t\!}d_{J_i} \: \Theta \: c^{}_{I_j})
\frac{\mu_j \: \epsilon_p^{I_j}}
{\nu_i \: \epsilon_p^{-J_i} - \mu_j \: \epsilon_p^{I_j}},
\]
and so, as well as the matrices $X_{\alpha, \cdots}$, also the
$\wt X_{\alpha, \cdots}$ can easily be written with the help of
(\ref{e:5.6}) just in form of $I_{n_\alpha}$ minus certain linear
combinations of $\wt Y^{(J)}_{\alpha,i j}$ and
$\wt Y^{(K)}_{\alpha,i j}$.

With the help of the relations (\ref{e:5.1}) and (\ref{e:5.5}),
(\ref{e:5.6}) it can be seen that
\[
C'_{-\alpha} = h^{-1}_{\alpha+1,J} C_{-\alpha} h^{}_{\alpha,J}
= C_{-\alpha},
\qquad
C'_{+\alpha} = h^{-1}_{\alpha,J} C_{+\alpha} h^{}_{\alpha+1,J}
= C_{+\alpha}.
\]
Hence, again, as it was the case for the one-soliton solutions,
the transformation
\[
h_{\alpha,J} \Gamma_\alpha \to \Gamma_\alpha
\]
is a symmetry transformation of the Toda equations (\ref{e:3.5}).
Consequently, we can write the multi-soliton solution to the
nonlinear matrix differential Toda equations (\ref{e:3.5})
as the `ratio'
\[
\Gamma_\alpha = {\wt T^{-1}_\alpha}{\wt T^{X}_\alpha},
\]
where the `numerator' and `denominator' are given explicitly
by (\ref{e:5.7}). Observe also that for $p=n$ one has $n_\alpha=1$
and then $\wt X_{\alpha,i_1 i_2 \ldots i_\ell}$ turns into
$\epsilon_p^{(\rho_{i_1} + \rho_{i_2} + \ldots + \rho_{i_\ell})}$,
and so, one obtains $\wt T^X_\alpha \to \wt T_{\alpha+1}$, with
$\wt T_\alpha$ reproducing the Hirota's $\tau$-functions for abelian
Toda systems \cite{NirRaz08a}.

It is also useful to have an explicit expression for the inverse
mapping, that is
\[
\Gamma^{-1}_{\alpha} = \wt T^{X^{-1}}_{\alpha+1} \:
\wt T^{-1}_{\alpha+1}.
\]
Here, the entries of this expression are defined according
to the relations (\ref{e:5.7}), with $\wt X_{\alpha,\ldots}$
replaced by
\[
\wt X^{-1}_{\alpha,i_1 i_2 \ldots i_\ell}
= X^{-1}_{\alpha,i_1 i_2 \ldots i_\ell} h^{}_{\alpha,J},
\]
where for $\ell = 1$ we have
\[
X^{-1}_{\alpha,i} = I_{n_\alpha} + \frac{1}{\wt F^{}_{i i}}
\sum^{r}_{j,k=1} \left(
\wt B^{-1}_{k j}(J) \: \wt F^{}_{i i}
- \wt B^{-1}_{k i}(J) \: \wt F^{}_{i j}
\right) \wt Y^{(J)}_{\alpha,j k}
+ \frac{1}{\wt F^{}_{i i}}
\sum^{r}_{k=1} \wt B^{-1}_{k i}(J)
\: \wt Y^{(K)}_{\alpha,i k},
\]
with the quantities $\wt F$ defined similarly to $\wt H$, only that
by means of $\wt B$,
\[
\wt F_{i j}(K,J)
= \sum_{k=1}^{r} \wt B^{}_{i k}(K) \: \wt B^{-1}_{k j}(J),
\]
and for $\ell \ge 2$ the other $n_\alpha \times n_\alpha$
inverse matrices $X^{-1}_{\alpha,\ldots}$ are
\begin{eqnarray*}
X^{-1}_{\alpha,i_1 i_2 \ldots i_\ell} &=& I_{n_\alpha}
+ \frac{1}{\displaystyle \sum_{\sigma \in S_\ell} \mathrm{sgn}(\sigma)
\displaystyle \prod^\ell_{m=1} \wt F^{}_{i_m i_{\sigma(m)}}}
\times \\
&& \times \left( \sum^{r}_{j,k=1}
\sum_{\sigma \in S'_{\ell+1}}
\mathrm{sgn}(\sigma) \:
\wt B^{-1}_{k \sigma(j)}(J) \:
\wt F^{}_{i_1 i_{\sigma(1)}}
\wt F^{}_{i_2 i_{\sigma(2)}} \ldots
\wt F^{}_{i_\ell i_{\sigma(\ell)}} \:
\wt Y^{(J)}_{\alpha,j k} \right. \\
&& \hskip5mm
+ \left. \sum^{r}_{k=1} \sum_{\sigma \in S_\ell}
\mathrm{sgn}(\sigma) \:
\wt B^{-1}_{k i_{\sigma(1)}}(J) \:
\wt F^{}_{i_2 i_{\sigma(2)}}
\wt F^{}_{i_3 i_{\sigma(3)}} \ldots
\wt F^{}_{i_\ell i_{\sigma(\ell)}} \:
\wt Y^{(K)}_{\alpha, i_1 k} \right).
\end{eqnarray*}

Now, to make our construction a little bit more transparent, we
add the following observation. Let us consider an $r \times r$
matrix $\wt \Delta_{i}$ being explicitly of the form
\[
\wt \Delta^{}_{i} =
\left(
\begin{array}{cccc}
  \wt D^{}_{11}(J) & \wt D^{}_{12}(J) & \ldots & \wt D^{}_{1r}(J) \\
       \vdots      &      \vdots      & \ddots & \vdots \\
\wt D^{}_{i,1}(K) & \wt D^{}_{i,2}(K) & \ldots & \wt D^{}_{i,r}(K) \\
       \vdots      &      \vdots      & \ddots & \vdots \\
  \wt D^{}_{r1}(J) & \wt D^{}_{r2}(J) & \ldots &
\wt D^{}_{rr}(J)
\end{array}
\right).
\]
That is, one takes the matrix $\wt D(J)$ and simply changes its $i$th
row by the corresponding row of the matrix $\wt D(K)$. Then we can write
\[
X_{\alpha,i} = I_{n_\alpha} - \sum^{r}_{\scriptstyle j,k=1
\atop \scriptstyle j \ne i}
(\wt \Delta^{-1}_{i})^{}_{k j}
\: \wt Y^{(J)}_{\alpha,j k}
- \sum^{r}_{k=1}
(\wt \Delta^{-1}_{i})^{}_{k i}
\: \wt Y^{(K)}_{\alpha,i k}.
\]
And, besides, one also has
\[
\rme^{\wt \delta_i} = \wt H^{}_{i i}
= \frac{\det \wt \Delta_{i}}{\det \wt D(J)}.
\]
In general, for $\ell \ge 2$, introduce the following
$r \times r$ matrix:
\[
\wt \Delta^{}_{i_1 i_2 \ldots i_\ell} =
\left(
\begin{array}{cccc}
  \wt D^{}_{11}(J) & \wt D^{}_{12}(J) & \ldots & \wt D^{}_{1r}(J) \\
       \vdots      &      \vdots      & \ddots & \vdots \\
\wt D^{}_{i_1 1}(K) & \wt D^{}_{i_1 2}(K) & \ldots &
\wt D^{}_{i_1 r}(K) \\
       \vdots      &      \vdots      & \ddots & \vdots \\
\wt D^{}_{i_2 1}(K) & \wt D^{}_{i_2 2}(K) & \ldots &
\wt D^{}_{i_2 r}(K) \\
       \vdots      &      \vdots      & \ddots & \vdots \\
\wt D^{}_{i_\ell 1}(K) & \wt D^{}_{i_\ell 2}(K) & \ldots &
\wt D^{}_{i_\ell r}(K) \\
       \vdots      &      \vdots      & \ddots & \vdots \\
  \wt D^{}_{r1}(J) & \wt D^{}_{r2}(J) & \ldots &
\wt D^{}_{rr}(J)
\end{array}
\right).
\]
Here, similarly to the preceding, we have taken the matrix
$\wt D(J)$ and replaced its rows $\wt D_{i_1 j}$,
$\wt D_{i_2 j}$, $\ldots$, $\wt D_{i_\ell j}$,
for $j$ running from $1$ to $r$, by the corresponding
matrix elements of $\wt D(K)$. And with the help of the
introduced matrices we immediately find out that
\[
X_{\alpha,i_1 i_2 \ldots i_\ell} = I_{n_\alpha} \: \: \: \: \: \:
- \sum^{r}_{\scriptstyle j,k=1 \atop \scriptstyle
j \ne i_1,i_2,\ldots,i_\ell}
(\wt \Delta^{-1}_{i_1 i_2 \ldots i_\ell})^{}_{k j}
\: \wt Y^{(J)}_{\alpha,j k} \: \: \: \: \: \: \: \: \: \:
- \sum^{r}_{\scriptstyle j,k=1 \atop \scriptstyle
j = i_1, i_2, \ldots, i_\ell}
(\wt \Delta^{-1}_{i_1 i_2 \ldots i_\ell})^{}_{k j}
\: \wt Y^{(K)}_{\alpha,j k}.
\]
Also the corresponding inverse matrices $X^{-1}_{\alpha,\ldots}$
can easily be found by replacing $\wt D$ by $\wt B$ in the above
construction and putting there the sign $+$ instead of $-$.

Note finally that if $n_* = 1$ some relations simplify, and we
recover for them certain expressions specific to the abelian
case \cite{NirRaz08a}.

\section{Conclusion}

In this paper we have considered the non-abelian Toda systems
associated with the untwisted loop groups of the complex general
linear groups. Developing the rational dressing method, we have
constructed multi-soliton solutions for these equations. Here,
the block-matrix representation of the group and algebra elements,
as suggested by the $\bbZ$-gradation, turned out to be most
appropriate to the problem under consideration. The solutions
are presented in a form of a direct matrix generalization of
the expressions obtained earlier for the abelian case
\cite{NirRaz08a}. In particular, the fact of non-abelian
generalization shows up explicitly through the special
matrices $\wt X_{\alpha,\ldots}$ and non-factorability of the
`soliton interaction coefficients' $\wt \eta_{i_1i_2\ldots}$.

We have also observed that the reduction to the non-abelian loop
Toda systems associated with the complex special linear groups can
be performed.

Now, it would be interesting to generalize our consideration to
other non-abelian loop Toda systems described in the classification
of \cite{NirRaz07a, NirRaz07b}, that is to Toda systems associated
with twisted loop groups of general linear groups and twisted and
untwisted loop groups of the complex orthogonal and symplectic Lie
groups.

\vskip2mm
{\bf Acknowledgments}

We are very grateful to the members of the Theoretical Physics Department 
of the University of Wuppertal, especially to Profs. Hermann Boos, Frank 
G\"ohmann and Andreas Kl\"umper for hospitality, certain interest to our 
work and stimulating discussions. This work was supported in part by the 
Russian Foundation for Basic Research under grant \#07--01--00234 and by 
the joint DFG--RFBR grant \#08--01--91953.

\end{document}